\documentclass[amsmath,amssymb,aps,twocolumn,superscriptaddress]{revtex4-2}
\usepackage{caption}
\usepackage[utf8]{inputenc}
\usepackage{graphicx}% Include figure files
\usepackage{dcolumn}% Align table columns on decimal point
\usepackage{bm}% bold math
\usepackage{siunitx}% try to write down numbers and units in a consistent way
\usepackage[version=4]{mhchem}% change $\mathrm{SO_2}$ or SiO${_2}$ to -> \ce{SiO2}
\usepackage{natbib}%customising references
\usepackage{float}%fix SI figures with specifier H
\usepackage[shortlabels]{enumitem}

\usepackage{xcolor}

\usepackage{mathtools}
\usepackage{braket}
\usepackage{amsmath}%
\usepackage{MnSymbol}%
\usepackage{wasysym}%
\usepackage{cancel}
\usepackage{physics}
\usepackage{caption}
\usepackage{subcaption}
\usepackage[english]{babel}
\usepackage[T1]{fontenc}
\usepackage{placeins}

\usepackage[colorlinks,citecolor=blue,urlcolor=red]{hyperref}
\usepackage{cleveref} 
\usepackage{chemformula}
\usepackage[newcommands]{ragged2e}

\makeatletter
\def\maketitle{
\@author@finish
\title@column\titleblock@produce
\suppressfloats[t]}
\makeatother
\newcommand{\beginsupplement}{
    \setcounter{table}{0}
    \renewcommand{\thetable}{S\arabic{table}}
    \setcounter{figure}{0}
    \renewcommand{\thefigure}{S\arabic{figure}}
    \setcounter{equation}{0}
    \setcounter{section}{0}
    \renewcommand{\theequation}{S\arabic{equation}}
}

\begin{document}
\title{Coherence measurements of polaritons in thermal equilibrium reveal a power law for two-dimensional condensates}

\author{Hassan Alnatah}
\thanks{Address correspondence to: haa108@pitt.edu}
\affiliation{Department of Physics, University of Pittsburgh, 3941 O’Hara Street, Pittsburgh, Pennsylvania 15218, USA}

\author{Qi Yao}
\affiliation{Department of Physics, University of Pittsburgh, 3941 O’Hara Street, Pittsburgh, Pennsylvania 15218, USA}

\author{Jonathan Beaumariage}
\affiliation{Department of Physics, University of Pittsburgh, 3941 O’Hara Street, Pittsburgh, Pennsylvania 15218, USA}

\author{Shouvik Mukherjee}
\affiliation{Joint Quantum Institute, University of Maryland and National Institute of Standards and Technology, College Park,
Maryland 20742, USA}

\author{Man Chun Tam}
\affiliation{Department of Electrical and Computer Engineering, University of Waterloo, Waterloo, ON, Canada}
\affiliation{Waterloo Institute for Nanotechnology, University of Waterloo, Waterloo, ON, Canada}

\author{Zbigniew Wasilewski}
\affiliation{Department of Electrical and Computer Engineering, University of Waterloo, Waterloo, ON, Canada}
\affiliation{Waterloo Institute for Nanotechnology, University of Waterloo, Waterloo, ON, Canada}

\author{Ken West}
\affiliation{Department of Electrical Engineering, Princeton University, Princeton, New Jersey 08544, USA}

\author{Kirk Baldwin}
\affiliation{Department of Electrical Engineering, Princeton University, Princeton, New Jersey 08544, USA}

\author{Loren N. Pfeiffer}
\affiliation{Department of Electrical Engineering, Princeton University, Princeton, New Jersey 08544, USA}

\author{David W. Snoke}
\affiliation{Department of Physics, University of Pittsburgh, 3941 O’Hara Street, Pittsburgh, Pennsylvania 15218, USA}

\date{\today}

\begin{abstract}
We have created a spatially homogeneous polariton condensate in thermal equilibrium, up to very high condensate fraction. Under these conditions, we have measured the coherence as a function of momentum, and determined the total coherent fraction of this boson system from very low density up to density well above the condensation transition.  These measurements reveal a consistent power law for the coherent fraction as a function of the total density over nearly three orders of its magnitude. The same power law is seen in numerical simulations solving the two-dimensional Gross-Pitaevskii equation for the equilibrium coherence. This power law has not been predicted by prior analytical theories.

\end{abstract}

\maketitle
%%%%%%%%%%%%%%%%%%%%%%%%%%%%%%%%%%%%%%%%%%%%%%%%%%%%%%
%%%%%%%%%%%%%%%%%%%%%%%%%%%%%%%%%%%%%%%%%%%%%%%%%%%%%%
\section{INTRODUCTION}
\par
\noindent
Bose-Einstein condensation (BEC) is a remarkable state of matter in which a macroscopically large number of bosons act as a single, coherent wave. The physics of two-dimensional BEC has subtle differences from the three-dimensional case because thermal fluctuations destroy the long-range order in systems of reduced dimensions \cite{mermin1966absence}. However, a quasicondensate state can exist with strongly correlated coherence over finite distances, as predicted by Berezinskii, Kosterlitz and Thouless \cite{berezinskii1971destruction,kosterlitz1973ordering}. 
Microcavity exciton-polaritons (called here simply ``polaritons'') are good candidates for the investigation of two-dimensional boson systems because they allow direct experimental accessibility to the coherence {\em in situ} without  destructive measurements. In most experiments with cold atoms that have tried to establish a phase diagram, only the momentum distribution or spatial profile has been measured, not the coherence directly \cite{plisson2011coherence,hadzibabic2006berezinskii,clade2009observation,navon2021quantum}.
\par 
%\red{I believe we should say a few words here about recent works on coherence properties of polaritons as well as citing Kavokin's paper. I have found several theory papers in the literature that are related to coherence of polaritons, for examples these theory papers \cite{malpuech2003polariton,whittaker2009coherence,doan2005condensation} and these experimental papers \cite{marchetti2008phase,pieczarka2022bogoliubov}. Most experimental papers I have seen assign the ground state occupation $n_{0}$ to be the coherent part. This is also true for the theory papers. It's either using a Quantum Boltzmann approach or Landau theory for the superfluid fraction. We could move this paragraph to the end of the introduction where we define what we mean by the coherent fraction (which is different than these papers define it) or keep it here.
%}
\par
Polaritons can be viewed as photons dressed with an effective mass and repulsive interactions, due to the strong coupling of a cavity photon state and a semiconductor exciton state. These particles have been shown to demonstrate Bose condensation and coherent effects in various experiments for nearly two decades (e.g. \cite{deng2002condensation,kasprzak2006bose,balili2007bose,abbarchi2013macroscopic,sanvitto2010persistent,lagoudakis2009observation}).
Although in many experiments the polaritons have fairly short lifetime, leading to nonequilibrium condensates, in the last ten years, microcavity structures have been available with polariton lifetime of several hundred picoseconds \cite{nelsen2013dissipationless,steger2015slow}, which has allowed demonstration of true equilibrium, as seen in near-perfect fits to an equilibrium Bose-Einstein energy distribution up to the Bose-degenerate regime \cite{sun2017bose} and in a thermal power law of the spatial correlation \cite{caputo2018topological}.

Ref.~\cite{sun2017bose} showed equilibrium in the degenerate regime up to 5-6 particles in the ground state, but at higher densities, the occupation-number distribution $N(E)$ deviated from a purely equilibrium distribution. We have since established that this was primarily due to the condensate becoming spatially inhomogeneous. In this work, we report new experiments in which equilibrium is well established in a homogeneous polariton gas well up to ground-state occupations in the range of 100-1000. Although the particles have a lifetime for decay which is replenished by a steady-state pump, the lifetime of the particles is long compared to their thermalization time, so that only a tiny fraction of the population is lost and replaced at any point in time.

This allows us to perform accurate measurements of the coherence of the gas over a wide range of density. Because the gas is thermal and homogeneous, it allows direct comparison to theories for the coherence of a Bose gas in two dimensions (2D). Although this type of experiment has been attempted with cold atoms \cite{chomaz2015emergence}, interference measurements in a cold atom gas are intrinsically a destructive measurement, and those measurements had low resolution. 

These experiments can be interpreted as measuring the ``condensate fraction'' of the system, but in a 2D system, the definition of the condensate fraction is somewhat controversial. Several theoretical papers (e.g., Refs. \cite{malpuech2003polariton, doan2005condensation}) have defined the ``condensate'' as only those particles with strictly zero momentum. These theories kept no track of the phase coherence, only the populations of $k$-states. However, the crucial aspect of the Gross-Pitaevskii equation, which allows superfluid behavior such as quantized vorticity, is the phase coherence, and the Gross-Pitaevskii equation makes no sharp distinction between the Fourier components of a coherent wave with $k=0$ and Fourier components with nonzero $k$. Also, if the condensate is defined as only particles with strictly zero momentum, then the condensate fraction in 2D has vanishingly small value in the thermodynamic limit; this is an unhelpful definition for a finite system, because it is well known \cite{mermin1966absence} that a 2D system can have coherence on finite length scales.  Instead, one can define the ``coherent fraction'' as the fraction of the particles which have 100\% fringe visibility in an interference measurement, which is equivalent to the integral of the fringe visibility over the total set of momentum states. This is well defined for any finite area. This may be termed the ``quasicondensate,'' since it corresponds to that part of the gas which has a single-valued wave function that obeys the Gross-Pitaevskii equation. Some, however, may restrict the term ``quasicondensate'' to a state with long-range, power-law correlation \cite{prokof2002two}, while at low density there is exponential decay of the correlation of the coherence (as seen in our numerical model, and presented in the Supplementary Information). What we see experimentally is that there is no sharp cutoff between the ``coherent fraction'' at low density and the ``quasicondensate'' at high density--the coherence increases continuously from very low density up to near 100\% at high density. In this paper, we will use the term ``coherent fraction'' to avoid confusion.
\par
In this work, we undertake a detailed experimental and theoretical investigation of coherence as a function of the polariton gas density and determine the coherent fraction as a function of total particle density. First, we establish the polariton gas is in thermal equilibrium. We then determine the coherent fraction and compare it to numerical solutions of a two-dimensional Gross-Pitaevskii equation. 
\section{EXPERIMENTAL OBSERVATIONS}
The microcavities used in this work consisted of a total of 12 GaAs quantum wells with AlAs barriers embedded within a distributed Bragg reflector (DBR). The DBRs are made of alternating layers of AlAs and \ce{Al_{0.2}Ga_{0.8}As}. The quantum wells are in groups of 4, with each group placed at one of the three antinodes of the $3\lambda/2$ cavity. The large number of DBR periods gives the cavity a high Q-factor, resulting in a cavity lifetime of $\sim$135 ps and a polariton life time of $\sim$270 ps at resonance. The long cavity life time allows polaritons to propagate over macroscopic distances of up to millimeters \cite{steger2015slow}. Further details about the samples are discussed in the Supplementary Information.
\par
The sample was cooled in a continuous-flow cold-finger cryostat at $\sim$5 K and excited non-resonantly with a continuous-wave (cw) laser, which was modulated by an optical chopper at 404 Hz with a duty cycle of 1.7\% to prevent sample heating. The pump profile was shaped into a broad Gaussian with full width at half maximum of $\sim 65 ~\mathrm{\mu m}$. The non-resonant excitation created a plasma of electrons and holes, which spontaneously form excitons. These hot excitons then scatter down in energy to become polaritons. 
\begin{figure}
\includegraphics[width=\columnwidth]{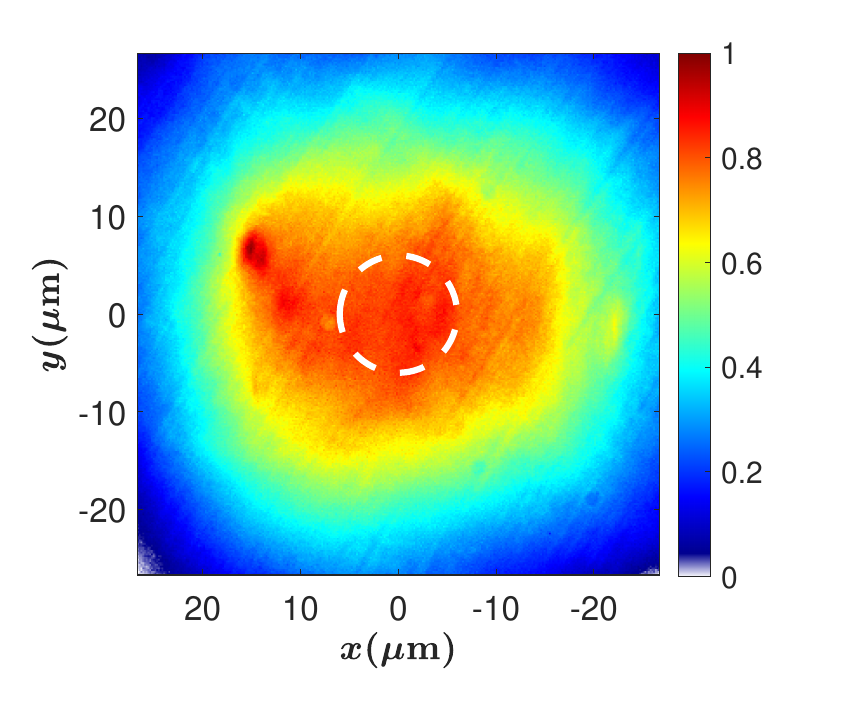}
\centering
\caption{\label{fig1} \textbf{Real-space polariton emission.} Polariton emission created by a wide area nonresonant pump. The white dashed circle indicates the region where the photoluminescence (PL) is collected.  \label{real_spce_PL}}
\end{figure}
\par
The cavity detuning was $\delta = 2.5 \; \text{meV}$, corresponding to an exciton fraction $\left | X \right |^{2} = 0.55$ for the lower polariton at $k=0$. The photoluminescence (PL) was collected using a microscope objective with a numerical aperture (NA) of 0.75, and was imaged onto the entrance slit of a spectrometer. The image was then sent through the spectrometer to a charged coupled device (CCD) for time integrated imaging. A spatial filter was placed at the real-space plane to collect PL from a region where the gas was very homogeneous (typical diameter of 12 $\mathrm{\mu m}$, as shown by the white dashed circle in Fig.~\ref{fig1}).

\begin{figure}
\includegraphics[width=\columnwidth]{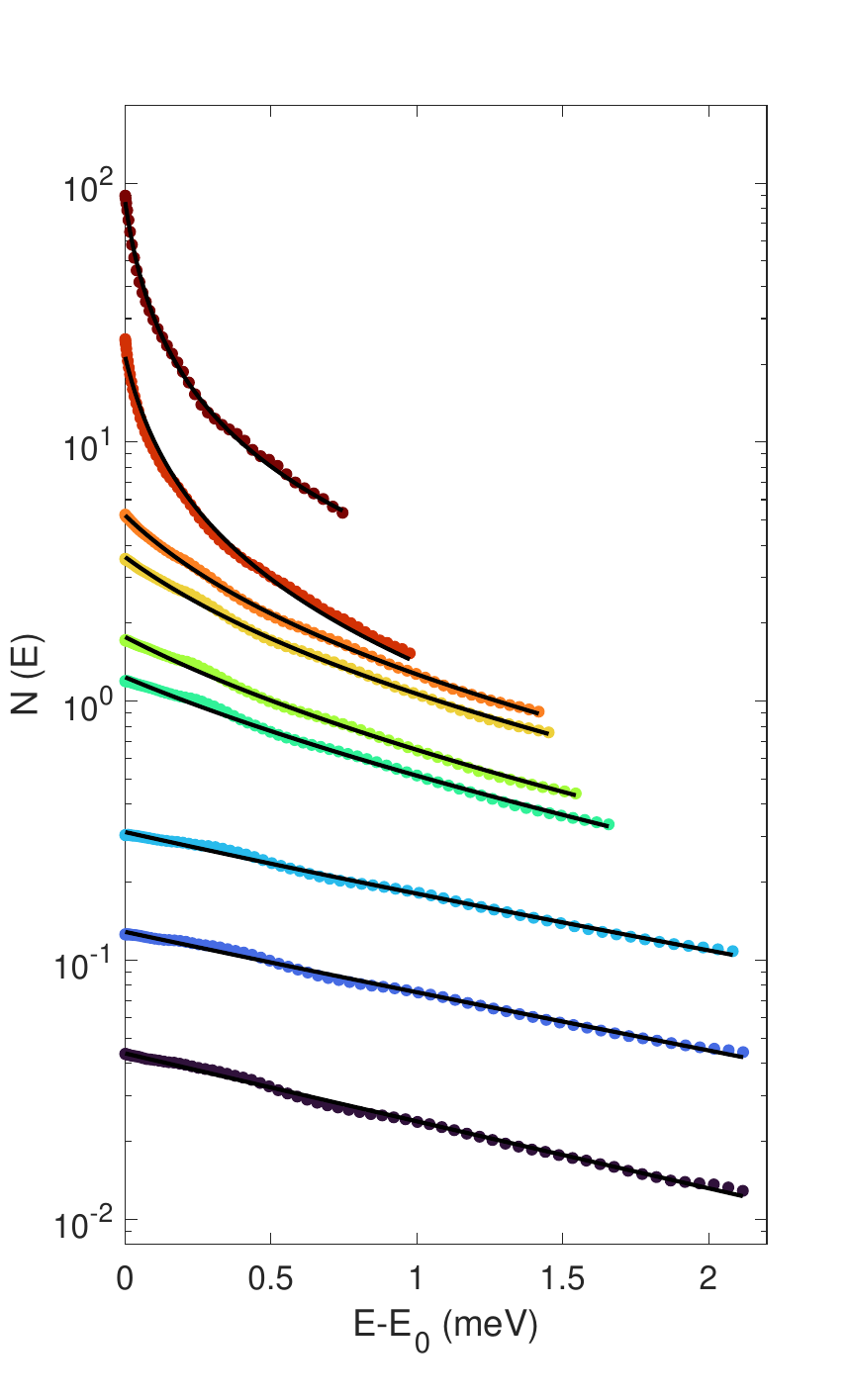}
\centering
\caption{\label{fig2} \textbf{Equilibrium distribution of polaritons}. Occupation of the lower polariton as a function of energy. The solid lines are best fits to the equilibrium Bose-Einstein distribution in Eq.~\eqref{eq:BE}. The temperature and chemical potential extracted from the fit are shown in Fig.~\ref{fig3}. The power values from low to high are 0.008, 0.031, 0.132, 0.530, 0.653, 0.821, 0.940, 1.164 and 1.265 times the threshold pump power. The threshold power $P_{\mathrm{th}}$ is defined in the Supplementary Information. }
\end{figure}
\par
\begin{figure}
\includegraphics[width=0.4\textwidth]{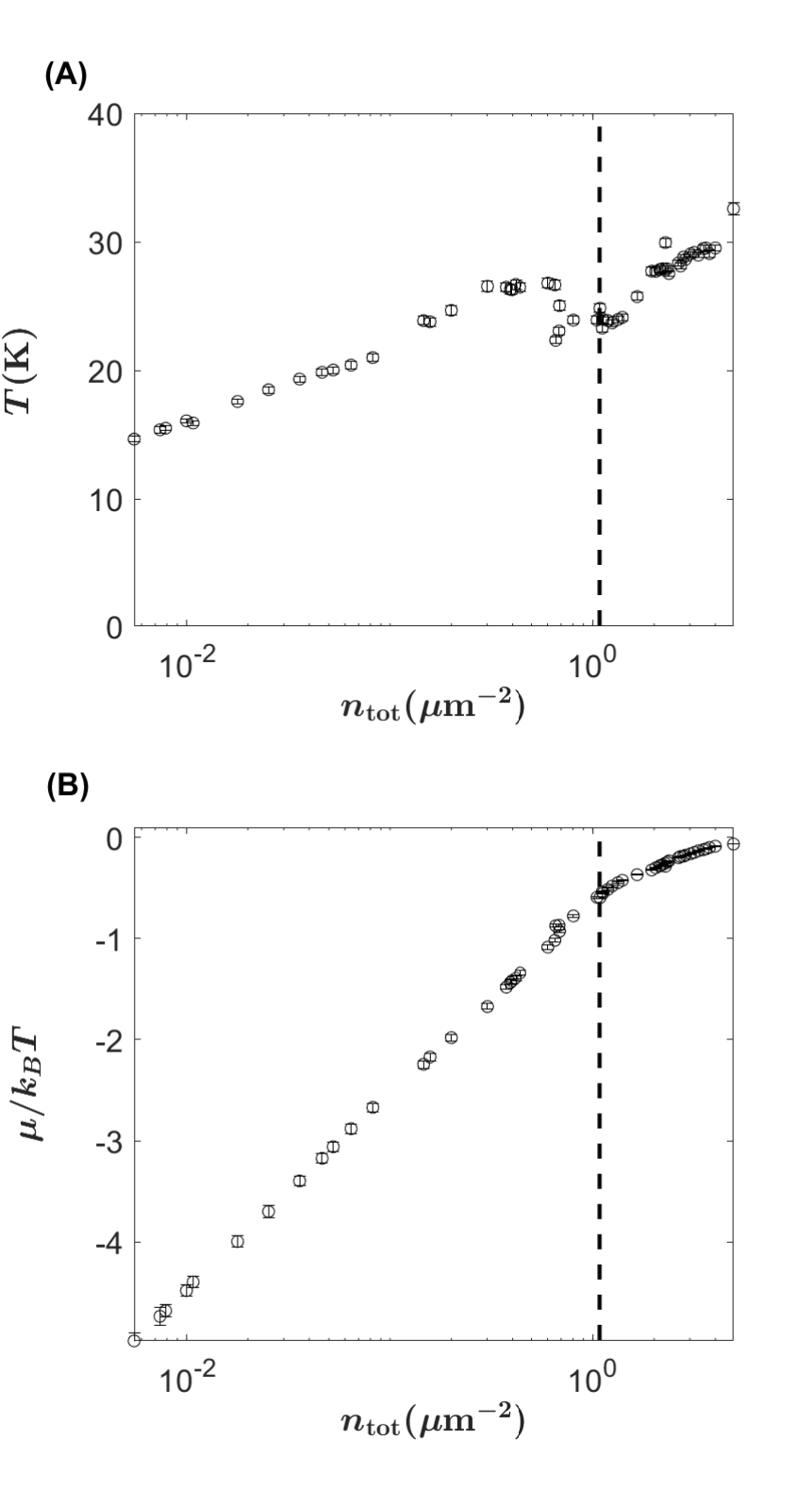}
\centering

\caption{\label{fig3} \textbf{Extracted temperature and chemical potential.} \textbf{(A)} The effective temperature of the polariton gas and \textbf{(B)} the reduced chemical potential obtained from the fits to the Bose-Einstein distribution. The vertical dashed line denotes when the occupation at $E=0$ becomes equal to one, i.e $N(E=0)=1$.}
\end{figure}

To show that polaritons can achieve thermal equilibrium, we measured the lower polariton occupation and we compared it to occupation number predicted by Bose–Einstein statistics. The lower-polariton occupation was measured by angle-resolved imaging, giving the intensity  $I\left ( k,E \right )$, which is then converted to an occupation number $N\left ( E \right )$ using a single efficiency factor (the calibration of this factor is discussed in the Supplementary Information). The measured polariton occupation for different pump power values is shown in Fig.~\ref{fig2}. The measured occupation numbers were fit to a Bose-Einstein distribution, given by

\begin{equation}
\begin{aligned}
N\left ( E_{\text{LP}} \right ) = \frac{1}{e^{\left (E_{\text{LP}}-E_{LP}(0)-\mu  \right )/k_{B}T}-1},
\label{eq:BE}
\end{aligned}
\end{equation}
where $T$ and $\mu$ are the temperature and chemical potential of the polariton gas respectively, $E_{\text{LP}}$ is the lower polariton energy, and $E_{LP}(0)$ is the polariton ground state energy at $k  = 0$, which shifts to higher energy as the density increases, due to many-body renormalization \cite{sun2017bose}. The fits to Bose-Einstein distribution were done by using $T$ and $\mu$ as fit parameters. As seen in Fig.~\ref{fig2}, the experimental polariton occupation is well described by a Bose-Einstein distribution for all densities indicating that the polariton gas is in true thermodynamic equilibrium. At densities well below the condensation threshold, the Bose-Einstein distribution becomes a Maxwell–Boltzmann distribution $N\left ( E_{LP} \right )\sim e^{\mu/k_{b}T}e^{-E_{LP}/k_{B}T}$, which corresponds to a straight line on a semilog plot. However, when quantum statistics become important (i.e., $N\left ( E_{LP} \right )\sim 1$), the shape of the distribution changes and an upturn at in low-energy states appears. The temperature and the chemical potential obtained from the fit to Bose-Einstein distribution are shown in Fig.~\ref{fig3}. We emphasize that a single efficiency factor is used for all the distributions and only $T$ and $\mu$ were varied.

\par
\begin{figure}
\includegraphics[width=\columnwidth]{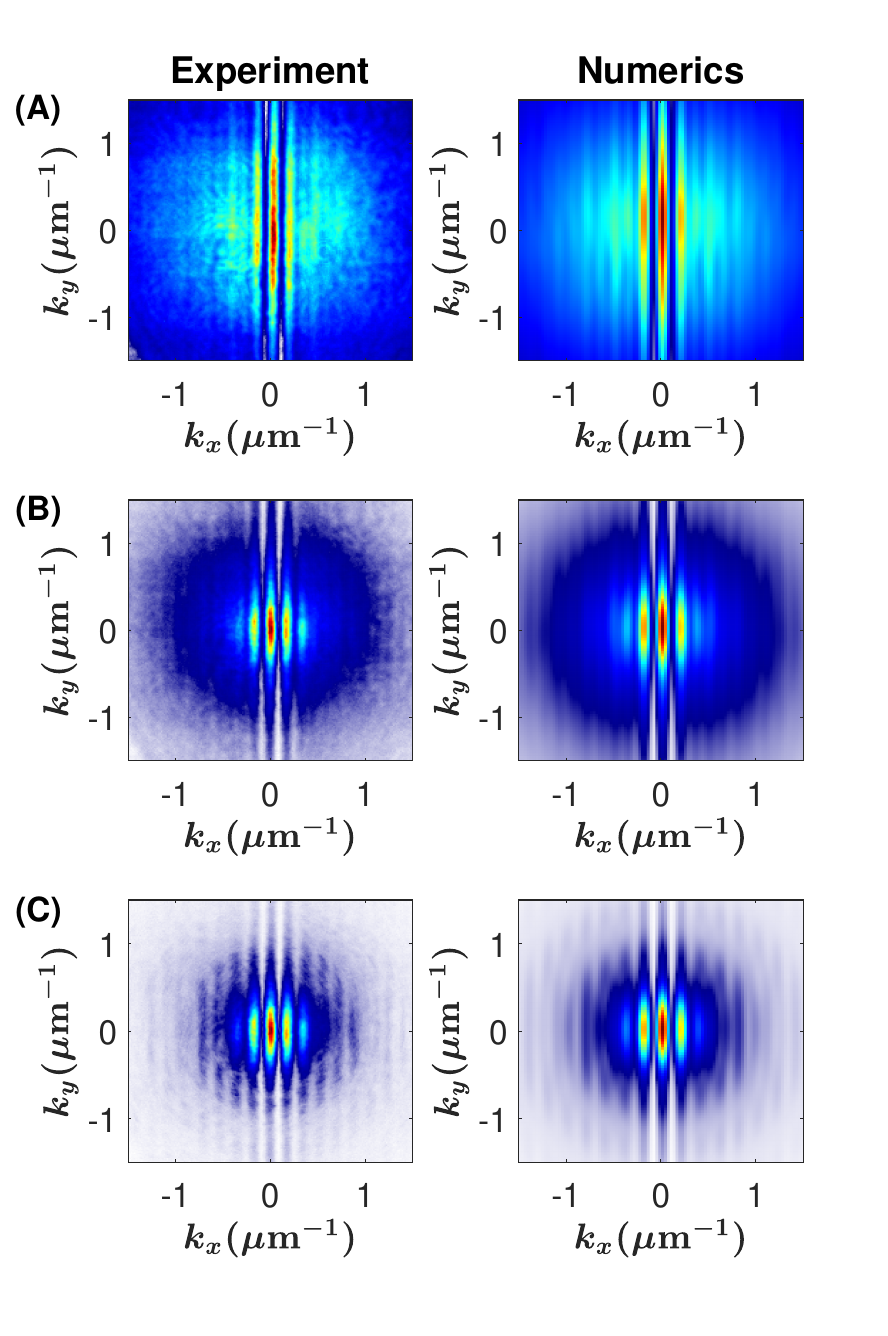}
\centering
\caption{\label{fig4}\textbf{Interference pattern.} The interference pattern in k-space obtained from the experiment (left column) and the numerics (right column) for three different densities, \textbf{(A)} $n = 1.5\;\mathrm{\mu m^{-2}}$, \textbf{(B)} $n = 4.5\;\mathrm{\mu m^{-2}}$ and \textbf{(C)} $n = 9.3\;\mathrm{\mu m^{-2}}$}
\end{figure}

The coherent fraction was measured by interfering the light emitted by the polariton gas $\vec{E}\left ( k_{x},k_{y},t_{0} \right )$ with its mirror symmetric image $\vec{E}\left (-k_{x},k_{y},t_{0} \right )$ using Michelson interferometry. The resulting intensity pattern exhibits interference fringes, indicating the emergence of extended coherence. A typical interference pattern in k-space is shown in Fig.~\ref{fig4} for different pump powers. We use these interference patterns to extract coherent fraction of the polariton gas as the fringe contrast gives a direct measurement of the level of coherence. 
\par
To extract the coherent fraction, we assume that interference pattern is described by a partially coherent wave with a momentum-dependent amplitude $N(k)$
\begin{equation}
\begin{aligned}
I (k)  =N(k)\left [ 1+ \alpha e^{-k/\kappa}\cos\left ( \lambda k \right )\right ],
\label{eq:interference_fit}
\end{aligned}
\end{equation}
where $\kappa$ is a fit parameter giving the region of coherence, $\alpha$ is a fit parameter ranging between 0 and 1 giving the degree of coherence, and $\lambda$ is the component associated with the fringe spacing. Therefore, the coherent fraction can be defined as:
\begin{equation}
\begin{aligned}
\frac{n_{0}}{n_{tot}}=\frac{\alpha\int \mathrm{d}^{2}k\; N(k)e^{-k/\kappa}}{ \int \mathrm{d}^{2}k\; N(k)}.
\label{eq:coherent_fraction}
\end{aligned}
\end{equation}
In the limit $\kappa \rightarrow  \infty$, Eq.~\eqref{eq:interference_fit} reduces to the interference pattern for fully coherent classical waves and the coherent fraction ${n_{0}}/{n_{tot}} \rightarrow 1$.

A striking result of these measurements is that the increase of the coherent fraction obeys a well-defined power law over a wide range of density, nearly three orders of its magnitude, as the density increases through the critical value. Figure~\ref{fig5} shows a typical data set; as shown in the Supplementary Information, many different data sets, including different values of the aperture size for the area of integration, can all be collapsed onto a single, universal curve. As discussed in the next section, this power law behavior is reproduced by a simple numerical solution of the Gross-Pitaevskii equation with no dissipation.
%%%%%%%%%%%%%%%%%%%%%%%%%%%%%%%%%%%%%%%%%%%%%%%%%%%%%%
\section{THEORY AND NUMERICAL SIMULATION}
\par
Because the experimental Bose gas is thermal and homogeneous, we can model the system using the Gross-Pitaevskii equation for the simplest case to get a universal result, which applies to any  number-conserving, spatially homogeneous, two-dimensional Bose gas in thermal equilibrium.
\par We solve the following Gross-Pitaevskii equation with noise introduced in the initial conditions, 
\begin{equation}
i \hbar \frac{\partial  \psi(\mathbf{r},t)}{\partial t}=\left[-\frac{\hbar^{2}\nabla^2 }{2m}+g_c|\psi(\mathbf{r},t)|^2\right] \psi(\mathbf{r},t),
\end{equation}
where $m$ is the mass of the polaritons and $g_{c}$ is the repulsive polariton-polariton interaction. Significantly, we do not include any terms for generation or decay of the polaritons, because as discussed above, the lifetime of the polaritons is long enough that these can be taken as negligible for the relevant dynamics, so that the system can be treated as number-conserving and in equilibrium.
\begin{figure}
\includegraphics[width=\columnwidth]{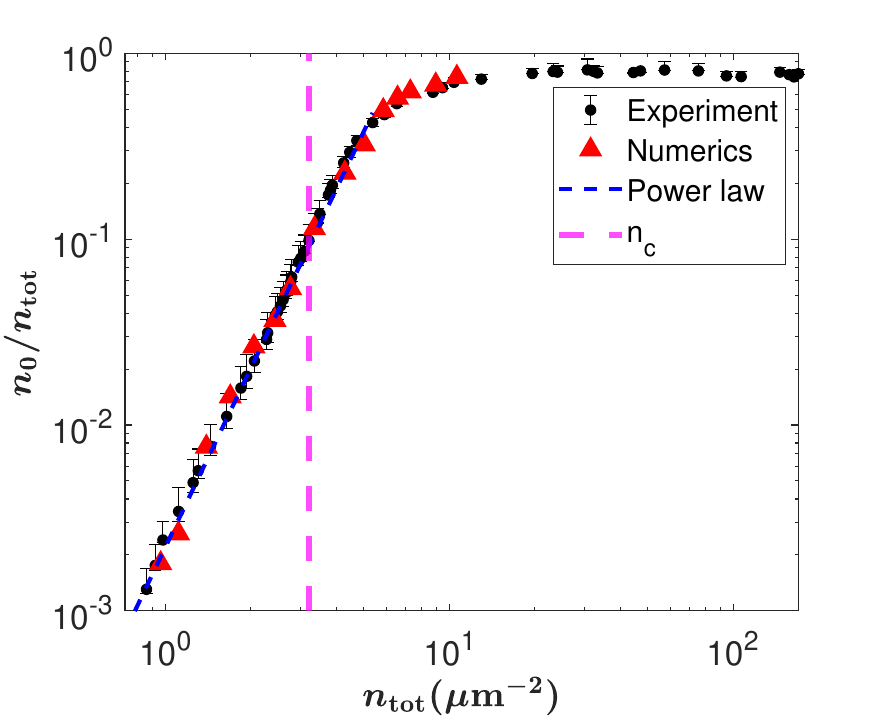}
\centering
\caption{\label{fig5} \textbf{Coherent fraction.} Black circles: experimentally measured coherent fraction as a function of the total polariton density for a pinhole with an area $A = \pi (6\;\si{\mu m})^{2}$. The quasicondensate fraction is defined in the text. Red triangles: coherent fraction defined the same way, for the numerical simulations. Blue line: $n^{3.2}$ power law. The vertical dashed line denotes the critical density, which is defined as the total density of polaritons at the threshold power $P/P_{\mathrm{th}} =1$, defined in the Supplementary Information.}
\end{figure}

\par To eliminate a computationally expensive transient regime, we start the system in an incoherent equilibrium state,
\begin{equation}
\begin{split}
\psi\left (x,y,t=0\right ) = \sum_{k_{n}}\sum_{k_{m}}\sqrt{ N\left (\sqrt{k^2_n+k^2_m} \right ) }\;e^{i\left ( k_{n}x+ k_{m}y\right )}\\
\times e^{i\left ( \theta_{k_{n}}+ \theta_{k_{m}}\right )},
\label{eq:initial_conditions}
\end{split}
\end{equation}
where $N\left (k  \right )  = (e^{(E(k)-\mu)/k_{B}T} -1)^{-1}$ is  the Bose-Einstein distribution and $E(k) = \hbar^2k^{2}/2m$. The phases $\theta_{k_{n}}$ and $\theta_{k_{m}}$ are random numbers that are uniformly distributed in the interval $\left [ 0,2\pi \right ]$. The system is then evolved in time until the system reaches a constant degree of coherence. To calculate the coherent fraction from the simulations, we first compute the interference pattern between $k_{x}$ and $-k_{x}$, 
\begin{equation}
\begin{split}
I\left ( k_{x},k_{y} \right )  =\frac{1}{t_{\mathrm{max}}}\int_{0}^{t_{\mathrm{max}}} \mathrm{d}t\; |\psi\left ( k_{x},k_{y},t \right )e^{ik_{x}x_{0}}\\
+\psi\left (-k_{x},k_{y},t \right )e^{-ik_{x}x_{0}} |^{2},
\end{split}
\end{equation}
where {$x_{0}$ is a constant that defines the fringe spacing and $t_{\mathrm{max}}$ is the total simulated time. $\psi\left ( k_{x},k_{y},t \right )$ is the Fourier transform of the wavefunction in real space. Since in the experiment we use a pinhole to only collect light from an area $A = \pi r^{2}$, we apply the same kind of filtering in the numerics for the real space wavefunction $\psi (x,y,t)$ before calculating the Fourier transform $\psi(k_{x},k_{y},t)$. The interference pattern $I\left ( k_{x},k_{y} \right )$ is evaluated by averaging over several independent stochastic paths for each random initial condition as described by Eq.~\eqref{eq:initial_conditions}. Figure \ref{fig4} shows a comparison between the experimentally measured interference pattern and the results obtained from the theoretical modeling, showing very good agreement for the fringe visibility for different densities. 
\section{CONCLUSIONS}
The coherent fraction from the numerical simulations is then calculated by following the same fitting procedure that was described in the previous section, namely Eqs.~(\ref{eq:interference_fit})-(\ref{eq:coherent_fraction}) (for more details, see the Supplementary Information). In both the experiment and the numerics, 
we subtracted the coherent fraction found in the zero-density limit, which corresponds to the coherence due to instrumental response, seen even in the Maxwell-Boltzmann limit. As seen in Figure~\ref{fig5}, our experimentally measured and numerically calculated coherent fraction show very good agreement, with the same $n^{3.2 \pm 0.12}$ power law over nearly three orders of magnitude of the value of the coherent fraction. At the highest densities, the coherent fraction of course cannot exceed unity, and therefore saturates. 

As discussed in the Supplementary Information, the numerical model also gives us the in-plane coherence length of the gas as a function of density, which gives the same power law of 3.2 when converted to an area. In general, the numerics allow us to explore a wide range of conditions, that agree with the experiments in all of the areas where we can compare them.

The agreement with the Gross-Pitaevskii numerical simulations for a homogeneous gas in equilibrium shows that the results of our experiments are truly universal, realizing the textbook paradigm of a uniform Bose Gas in two-dimensions in thermal equilibrium. 

\par Although the coherent fraction depends on the area from which the light is collected, we show in the Supplementary Information that the same power law is experimentally observed for different pinhole sizes. The largest pinhole that was used experimentally has a diameter of $12\; \si{\mu m}$ since for larger pinhole sizes, the assumption of homogeneity breaks down. However, our numerical model allows us to explore the effect of larger pinhole sizes. In agreement with the experiment, our numerical model shows that the effect of the aperture size gives a shifted curve with the same $n^{3.2}$ power law (see Supplementary Information). Of course, for an infinite system, the coherent fraction goes to zero since an infinite two-dimensional system cannot have long-range order at any finite temperature, but for any finite area of observation, the same power law will be valid.  

\par The density dependence of properties of a two-dimensional condensate has not been deeply explored in the literature, because typical experiments and theory assume a constant density and variation of temperature. We are not aware of any predictions of the observed power law for the coherent fraction, but since this appears in clearly in both the experiments and simulations for a thermal, homogeneous gas, this should be a universal result. We emphasize the need for further theoretical exploration to give more physical intuition into the origin of the observed power law. It is our hope that our findings will inspire additional theoretical research to understand more deeply this universal power law.

\par It is quite surprising that any new universal behaviors could be found in a field as well studied for the past 50 years as two-dimensional condensates.  This is made possible by the experimental advances of very fine control over the polariton density and long lifetime which allows equilibrium over a wide range of density, as well as the direct {\em in situ} measurement of coherence, which is not possible in liquid helium or cold atoms. 
\section{ACKNOWLEDGEMENTS}
We thank Boris Svistunov, Marzena Szymanska and Paolo Comaron for fruitful discussions.
\section{FUNDING}
The experimental work at Pittsburgh and sample fabrication at Princeton and Waterloo were supported by the National Science Foundation, Grant No. DMR-2004570. Optimization of epitaxial growth and material characterization at the University of Waterloo were partly funded by the Natural Sciences and Engineering Research Council of Canada, Grant No. RGPIN-05345-2018.
%%%%%%%%%%%%%%%%%%%%%%%%%%%%%%%%%%%%%%%%%
%%%%%%%%%%%%%%%%%%%%%%%%%%%%%%%%%%%%%%%%%%%%%%%%%%%%%%
%%%%%%%%%%%%%%%%%%%%%%%%%%%%%%%%%%%%%%%%%%%%%%%%%%%%%%
%%%%%%%%%%%%%%%%%%%%%%%%%%%%%%%%%%%%%%%%
\bibliography{references.bib}

\begin{thebibliography}{22}%
\makeatletter
\providecommand \@ifxundefined [1]{%
 \@ifx{#1\undefined}
}%
\providecommand \@ifnum [1]{%
 \ifnum #1\expandafter \@firstoftwo
 \else \expandafter \@secondoftwo
 \fi
}%
\providecommand \@ifx [1]{%
 \ifx #1\expandafter \@firstoftwo
 \else \expandafter \@secondoftwo
 \fi
}%
\providecommand \natexlab [1]{#1}%
\providecommand \enquote  [1]{``#1''}%
\providecommand \bibnamefont  [1]{#1}%
\providecommand \bibfnamefont [1]{#1}%
\providecommand \citenamefont [1]{#1}%
\providecommand \href@noop [0]{\@secondoftwo}%
\providecommand \href [0]{\begingroup \@sanitize@url \@href}%
\providecommand \@href[1]{\@@startlink{#1}\@@href}%
\providecommand \@@href[1]{\endgroup#1\@@endlink}%
\providecommand \@sanitize@url [0]{\catcode `\\12\catcode `\$12\catcode `\&12\catcode `\#12\catcode `\^12\catcode `\_12\catcode `\%12\relax}%
\providecommand \@@startlink[1]{}%
\providecommand \@@endlink[0]{}%
\providecommand \url  [0]{\begingroup\@sanitize@url \@url }%
\providecommand \@url [1]{\endgroup\@href {#1}{\urlprefix }}%
\providecommand \urlprefix  [0]{URL }%
\providecommand \Eprint [0]{\href }%
\providecommand \doibase [0]{https://doi.org/}%
\providecommand \selectlanguage [0]{\@gobble}%
\providecommand \bibinfo  [0]{\@secondoftwo}%
\providecommand \bibfield  [0]{\@secondoftwo}%
\providecommand \translation [1]{[#1]}%
\providecommand \BibitemOpen [0]{}%
\providecommand \bibitemStop [0]{}%
\providecommand \bibitemNoStop [0]{.\EOS\space}%
\providecommand \EOS [0]{\spacefactor3000\relax}%
\providecommand \BibitemShut  [1]{\csname bibitem#1\endcsname}%
\let\auto@bib@innerbib\@empty
%</preamble>
\bibitem [{\citenamefont {Mermin}\ and\ \citenamefont {Wagner}(1966)}]{mermin1966absence}%
  \BibitemOpen
  \bibfield  {author} {\bibinfo {author} {\bibfnamefont {N.~D.}\ \bibnamefont {Mermin}}\ and\ \bibinfo {author} {\bibfnamefont {H.}~\bibnamefont {Wagner}},\ }\bibfield  {title} {\bibinfo {title} {Absence of {F}erromagnetism or {A}ntiferromagnetism in {O}ne-or {T}wo-dimensional {I}sotropic {H}eisenberg {M}odels},\ }\href@noop {} {\bibfield  {journal} {\bibinfo  {journal} {Physical Review Letters}\ }\textbf {\bibinfo {volume} {17}},\ \bibinfo {pages} {1133} (\bibinfo {year} {1966})}\BibitemShut {NoStop}%
\bibitem [{\citenamefont {Berezinskii}(1971)}]{berezinskii1971destruction}%
  \BibitemOpen
  \bibfield  {author} {\bibinfo {author} {\bibfnamefont {V.}~\bibnamefont {Berezinskii}},\ }\bibfield  {title} {\bibinfo {title} {Destruction of {L}ong-range {O}rder in {O}ne-dimensional and {T}wo-dimensional {S}ystems {H}aving a {C}ontinuous {S}ymmetry {G}roup {I}. {C}lassical {S}ystems},\ }\href@noop {} {\bibfield  {journal} {\bibinfo  {journal} {Sov. Phys. JETP}\ }\textbf {\bibinfo {volume} {32}},\ \bibinfo {pages} {493} (\bibinfo {year} {1971})}\BibitemShut {NoStop}%
\bibitem [{\citenamefont {Kosterlitz}\ and\ \citenamefont {Thouless}(1973)}]{kosterlitz1973ordering}%
  \BibitemOpen
  \bibfield  {author} {\bibinfo {author} {\bibfnamefont {J.~M.}\ \bibnamefont {Kosterlitz}}\ and\ \bibinfo {author} {\bibfnamefont {D.~J.}\ \bibnamefont {Thouless}},\ }\bibfield  {title} {\bibinfo {title} {Ordering, {M}etastability and {P}hase {T}ransitions in {T}wo-dimensional {S}ystems},\ }\href@noop {} {\bibfield  {journal} {\bibinfo  {journal} {Journal of Physics C: Solid State Physics}\ }\textbf {\bibinfo {volume} {6}},\ \bibinfo {pages} {1181} (\bibinfo {year} {1973})}\BibitemShut {NoStop}%
\bibitem [{\citenamefont {Plisson}\ \emph {et~al.}(2011)\citenamefont {Plisson}, \citenamefont {Allard}, \citenamefont {Holzmann}, \citenamefont {Salomon}, \citenamefont {Aspect}, \citenamefont {Bouyer},\ and\ \citenamefont {Bourdel}}]{plisson2011coherence}%
  \BibitemOpen
  \bibfield  {author} {\bibinfo {author} {\bibfnamefont {T.}~\bibnamefont {Plisson}}, \bibinfo {author} {\bibfnamefont {B.}~\bibnamefont {Allard}}, \bibinfo {author} {\bibfnamefont {M.}~\bibnamefont {Holzmann}}, \bibinfo {author} {\bibfnamefont {G.}~\bibnamefont {Salomon}}, \bibinfo {author} {\bibfnamefont {A.}~\bibnamefont {Aspect}}, \bibinfo {author} {\bibfnamefont {P.}~\bibnamefont {Bouyer}},\ and\ \bibinfo {author} {\bibfnamefont {T.}~\bibnamefont {Bourdel}},\ }\bibfield  {title} {\bibinfo {title} {Coherence {P}roperties of a {T}wo-dimensional {T}rapped {B}ose {G}as {A}round the {S}uperfluid {T}ransition},\ }\href@noop {} {\bibfield  {journal} {\bibinfo  {journal} {Physical Review A}\ }\textbf {\bibinfo {volume} {84}},\ \bibinfo {pages} {061606} (\bibinfo {year} {2011})}\BibitemShut {NoStop}%
\bibitem [{\citenamefont {Hadzibabic}\ \emph {et~al.}(2006)\citenamefont {Hadzibabic}, \citenamefont {Kr{\"u}ger}, \citenamefont {Cheneau}, \citenamefont {Battelier},\ and\ \citenamefont {Dalibard}}]{hadzibabic2006berezinskii}%
  \BibitemOpen
  \bibfield  {author} {\bibinfo {author} {\bibfnamefont {Z.}~\bibnamefont {Hadzibabic}}, \bibinfo {author} {\bibfnamefont {P.}~\bibnamefont {Kr{\"u}ger}}, \bibinfo {author} {\bibfnamefont {M.}~\bibnamefont {Cheneau}}, \bibinfo {author} {\bibfnamefont {B.}~\bibnamefont {Battelier}},\ and\ \bibinfo {author} {\bibfnamefont {J.}~\bibnamefont {Dalibard}},\ }\bibfield  {title} {\bibinfo {title} {Berezinskii--{K}osterlitz--{T}houless {C}rossover in a {T}rapped {A}tomic {G}as},\ }\href@noop {} {\bibfield  {journal} {\bibinfo  {journal} {Nature}\ }\textbf {\bibinfo {volume} {441}},\ \bibinfo {pages} {1118} (\bibinfo {year} {2006})}\BibitemShut {NoStop}%
\bibitem [{\citenamefont {Clade}\ \emph {et~al.}(2009)\citenamefont {Clade}, \citenamefont {Ryu}, \citenamefont {Ramanathan}, \citenamefont {Helmerson},\ and\ \citenamefont {Phillips}}]{clade2009observation}%
  \BibitemOpen
  \bibfield  {author} {\bibinfo {author} {\bibfnamefont {P.}~\bibnamefont {Clade}}, \bibinfo {author} {\bibfnamefont {C.}~\bibnamefont {Ryu}}, \bibinfo {author} {\bibfnamefont {A.}~\bibnamefont {Ramanathan}}, \bibinfo {author} {\bibfnamefont {K.}~\bibnamefont {Helmerson}},\ and\ \bibinfo {author} {\bibfnamefont {W.~D.}\ \bibnamefont {Phillips}},\ }\bibfield  {title} {\bibinfo {title} {Observation of a 2{D} {B}ose {G}as: from {T}hermal to {Q}uasicondensate to {S}uperfluid},\ }\href@noop {} {\bibfield  {journal} {\bibinfo  {journal} {Physical review letters}\ }\textbf {\bibinfo {volume} {102}},\ \bibinfo {pages} {170401} (\bibinfo {year} {2009})}\BibitemShut {NoStop}%
\bibitem [{\citenamefont {Navon}\ \emph {et~al.}(2021)\citenamefont {Navon}, \citenamefont {Smith},\ and\ \citenamefont {Hadzibabic}}]{navon2021quantum}%
  \BibitemOpen
  \bibfield  {author} {\bibinfo {author} {\bibfnamefont {N.}~\bibnamefont {Navon}}, \bibinfo {author} {\bibfnamefont {R.~P.}\ \bibnamefont {Smith}},\ and\ \bibinfo {author} {\bibfnamefont {Z.}~\bibnamefont {Hadzibabic}},\ }\bibfield  {title} {\bibinfo {title} {Quantum {G}ases in {O}ptical {B}oxes},\ }\href@noop {} {\bibfield  {journal} {\bibinfo  {journal} {Nature Physics}\ }\textbf {\bibinfo {volume} {17}},\ \bibinfo {pages} {1334} (\bibinfo {year} {2021})}\BibitemShut {NoStop}%
\bibitem [{\citenamefont {Deng}\ \emph {et~al.}(2002)\citenamefont {Deng}, \citenamefont {Weihs}, \citenamefont {Santori}, \citenamefont {Bloch},\ and\ \citenamefont {Yamamoto}}]{deng2002condensation}%
  \BibitemOpen
  \bibfield  {author} {\bibinfo {author} {\bibfnamefont {H.}~\bibnamefont {Deng}}, \bibinfo {author} {\bibfnamefont {G.}~\bibnamefont {Weihs}}, \bibinfo {author} {\bibfnamefont {C.}~\bibnamefont {Santori}}, \bibinfo {author} {\bibfnamefont {J.}~\bibnamefont {Bloch}},\ and\ \bibinfo {author} {\bibfnamefont {Y.}~\bibnamefont {Yamamoto}},\ }\bibfield  {title} {\bibinfo {title} {Condensation of {S}emiconductor {M}icrocavity {E}xciton {P}olaritons},\ }\href@noop {} {\bibfield  {journal} {\bibinfo  {journal} {Science}\ }\textbf {\bibinfo {volume} {298}},\ \bibinfo {pages} {199} (\bibinfo {year} {2002})}\BibitemShut {NoStop}%
\bibitem [{\citenamefont {Kasprzak}\ \emph {et~al.}(2006)\citenamefont {Kasprzak}, \citenamefont {Richard}, \citenamefont {Kundermann}, \citenamefont {Baas}, \citenamefont {Jeambrun}, \citenamefont {Keeling}, \citenamefont {Marchetti}, \citenamefont {Szyma{\'n}ska}, \citenamefont {Andr{\'e}}, \citenamefont {Staehli} \emph {et~al.}}]{kasprzak2006bose}%
  \BibitemOpen
  \bibfield  {author} {\bibinfo {author} {\bibfnamefont {J.}~\bibnamefont {Kasprzak}}, \bibinfo {author} {\bibfnamefont {M.}~\bibnamefont {Richard}}, \bibinfo {author} {\bibfnamefont {S.}~\bibnamefont {Kundermann}}, \bibinfo {author} {\bibfnamefont {A.}~\bibnamefont {Baas}}, \bibinfo {author} {\bibfnamefont {P.}~\bibnamefont {Jeambrun}}, \bibinfo {author} {\bibfnamefont {J.~M.~J.}\ \bibnamefont {Keeling}}, \bibinfo {author} {\bibfnamefont {F.}~\bibnamefont {Marchetti}}, \bibinfo {author} {\bibfnamefont {M.}~\bibnamefont {Szyma{\'n}ska}}, \bibinfo {author} {\bibfnamefont {R.}~\bibnamefont {Andr{\'e}}}, \bibinfo {author} {\bibfnamefont {J.}~\bibnamefont {Staehli}}, \emph {et~al.},\ }\bibfield  {title} {\bibinfo {title} {Bose--{E}instein {C}ondensation of {E}xciton {P}olaritons},\ }\href@noop {} {\bibfield  {journal} {\bibinfo  {journal} {Nature}\ }\textbf {\bibinfo {volume} {443}},\ \bibinfo {pages} {409} (\bibinfo {year} {2006})}\BibitemShut {NoStop}%
\bibitem [{\citenamefont {Balili}\ \emph {et~al.}(2007)\citenamefont {Balili}, \citenamefont {Hartwell}, \citenamefont {Snoke}, \citenamefont {Pfeiffer},\ and\ \citenamefont {West}}]{balili2007bose}%
  \BibitemOpen
  \bibfield  {author} {\bibinfo {author} {\bibfnamefont {R.}~\bibnamefont {Balili}}, \bibinfo {author} {\bibfnamefont {V.}~\bibnamefont {Hartwell}}, \bibinfo {author} {\bibfnamefont {D.}~\bibnamefont {Snoke}}, \bibinfo {author} {\bibfnamefont {L.}~\bibnamefont {Pfeiffer}},\ and\ \bibinfo {author} {\bibfnamefont {K.}~\bibnamefont {West}},\ }\bibfield  {title} {\bibinfo {title} {Bose-{E}instein {C}ondensation of {M}icrocavity {P}olaritons in a {T}rap},\ }\href@noop {} {\bibfield  {journal} {\bibinfo  {journal} {Science}\ }\textbf {\bibinfo {volume} {316}},\ \bibinfo {pages} {1007} (\bibinfo {year} {2007})}\BibitemShut {NoStop}%
\bibitem [{\citenamefont {Abbarchi}\ \emph {et~al.}(2013)\citenamefont {Abbarchi}, \citenamefont {Amo}, \citenamefont {Sala}, \citenamefont {Solnyshkov}, \citenamefont {Flayac}, \citenamefont {Ferrier}, \citenamefont {Sagnes}, \citenamefont {Galopin}, \citenamefont {Lema{\^\i}tre}, \citenamefont {Malpuech} \emph {et~al.}}]{abbarchi2013macroscopic}%
  \BibitemOpen
  \bibfield  {author} {\bibinfo {author} {\bibfnamefont {M.}~\bibnamefont {Abbarchi}}, \bibinfo {author} {\bibfnamefont {A.}~\bibnamefont {Amo}}, \bibinfo {author} {\bibfnamefont {V.}~\bibnamefont {Sala}}, \bibinfo {author} {\bibfnamefont {D.}~\bibnamefont {Solnyshkov}}, \bibinfo {author} {\bibfnamefont {H.}~\bibnamefont {Flayac}}, \bibinfo {author} {\bibfnamefont {L.}~\bibnamefont {Ferrier}}, \bibinfo {author} {\bibfnamefont {I.}~\bibnamefont {Sagnes}}, \bibinfo {author} {\bibfnamefont {E.}~\bibnamefont {Galopin}}, \bibinfo {author} {\bibfnamefont {A.}~\bibnamefont {Lema{\^\i}tre}}, \bibinfo {author} {\bibfnamefont {G.}~\bibnamefont {Malpuech}}, \emph {et~al.},\ }\bibfield  {title} {\bibinfo {title} {Macroscopic {Q}uantum {S}elf-trapping and {J}osephson {O}scillations of {E}xciton {P}olaritons},\ }\href@noop {} {\bibfield  {journal} {\bibinfo  {journal} {Nature Physics}\ }\textbf {\bibinfo {volume} {9}},\ \bibinfo {pages} {275} (\bibinfo {year} {2013})}\BibitemShut {NoStop}%
\bibitem [{\citenamefont {Sanvitto}\ \emph {et~al.}(2010)\citenamefont {Sanvitto}, \citenamefont {Marchetti}, \citenamefont {Szyma{\'n}ska}, \citenamefont {Tosi}, \citenamefont {Baudisch}, \citenamefont {Laussy}, \citenamefont {Krizhanovskii}, \citenamefont {Skolnick}, \citenamefont {Marrucci}, \citenamefont {Lemaitre} \emph {et~al.}}]{sanvitto2010persistent}%
  \BibitemOpen
  \bibfield  {author} {\bibinfo {author} {\bibfnamefont {D.}~\bibnamefont {Sanvitto}}, \bibinfo {author} {\bibfnamefont {F.}~\bibnamefont {Marchetti}}, \bibinfo {author} {\bibfnamefont {M.}~\bibnamefont {Szyma{\'n}ska}}, \bibinfo {author} {\bibfnamefont {G.}~\bibnamefont {Tosi}}, \bibinfo {author} {\bibfnamefont {M.}~\bibnamefont {Baudisch}}, \bibinfo {author} {\bibfnamefont {F.~P.}\ \bibnamefont {Laussy}}, \bibinfo {author} {\bibfnamefont {D.}~\bibnamefont {Krizhanovskii}}, \bibinfo {author} {\bibfnamefont {M.}~\bibnamefont {Skolnick}}, \bibinfo {author} {\bibfnamefont {L.}~\bibnamefont {Marrucci}}, \bibinfo {author} {\bibfnamefont {A.}~\bibnamefont {Lemaitre}}, \emph {et~al.},\ }\bibfield  {title} {\bibinfo {title} {Persistent {C}urrents and {Q}uantized {V}ortices in a {P}olariton {S}uperfluid},\ }\href@noop {} {\bibfield  {journal} {\bibinfo  {journal} {Nature Physics}\ }\textbf {\bibinfo {volume} {6}},\ \bibinfo {pages} {527} (\bibinfo {year} {2010})}\BibitemShut {NoStop}%
\bibitem [{\citenamefont {Lagoudakis}\ \emph {et~al.}(2009)\citenamefont {Lagoudakis}, \citenamefont {Ostatnick{\`y}}, \citenamefont {Kavokin}, \citenamefont {Rubo}, \citenamefont {Andr{\'e}},\ and\ \citenamefont {Deveaud-Pl{\'e}dran}}]{lagoudakis2009observation}%
  \BibitemOpen
  \bibfield  {author} {\bibinfo {author} {\bibfnamefont {K.}~\bibnamefont {Lagoudakis}}, \bibinfo {author} {\bibfnamefont {T.}~\bibnamefont {Ostatnick{\`y}}}, \bibinfo {author} {\bibfnamefont {A.}~\bibnamefont {Kavokin}}, \bibinfo {author} {\bibfnamefont {Y.~G.}\ \bibnamefont {Rubo}}, \bibinfo {author} {\bibfnamefont {R.}~\bibnamefont {Andr{\'e}}},\ and\ \bibinfo {author} {\bibfnamefont {B.}~\bibnamefont {Deveaud-Pl{\'e}dran}},\ }\bibfield  {title} {\bibinfo {title} {Observation of {H}alf-quantum {V}ortices in an {E}xciton-polariton {C}ondensate},\ }\href@noop {} {\bibfield  {journal} {\bibinfo  {journal} {science}\ }\textbf {\bibinfo {volume} {326}},\ \bibinfo {pages} {974} (\bibinfo {year} {2009})}\BibitemShut {NoStop}%
\bibitem [{\citenamefont {Nelsen}\ \emph {et~al.}(2013)\citenamefont {Nelsen}, \citenamefont {Liu}, \citenamefont {Steger}, \citenamefont {Snoke}, \citenamefont {Balili}, \citenamefont {West},\ and\ \citenamefont {Pfeiffer}}]{nelsen2013dissipationless}%
  \BibitemOpen
  \bibfield  {author} {\bibinfo {author} {\bibfnamefont {B.}~\bibnamefont {Nelsen}}, \bibinfo {author} {\bibfnamefont {G.}~\bibnamefont {Liu}}, \bibinfo {author} {\bibfnamefont {M.}~\bibnamefont {Steger}}, \bibinfo {author} {\bibfnamefont {D.~W.}\ \bibnamefont {Snoke}}, \bibinfo {author} {\bibfnamefont {R.}~\bibnamefont {Balili}}, \bibinfo {author} {\bibfnamefont {K.}~\bibnamefont {West}},\ and\ \bibinfo {author} {\bibfnamefont {L.}~\bibnamefont {Pfeiffer}},\ }\bibfield  {title} {\bibinfo {title} {Dissipationless {F}low and {S}harp {T}hreshold of a {P}olariton {C}ondensate with {L}ong {L}ifetime},\ }\href@noop {} {\bibfield  {journal} {\bibinfo  {journal} {Physical Review X}\ }\textbf {\bibinfo {volume} {3}},\ \bibinfo {pages} {041015} (\bibinfo {year} {2013})}\BibitemShut {NoStop}%
\bibitem [{\citenamefont {Steger}\ \emph {et~al.}(2015)\citenamefont {Steger}, \citenamefont {Gautham}, \citenamefont {Snoke}, \citenamefont {Pfeiffer},\ and\ \citenamefont {West}}]{steger2015slow}%
  \BibitemOpen
  \bibfield  {author} {\bibinfo {author} {\bibfnamefont {M.}~\bibnamefont {Steger}}, \bibinfo {author} {\bibfnamefont {C.}~\bibnamefont {Gautham}}, \bibinfo {author} {\bibfnamefont {D.~W.}\ \bibnamefont {Snoke}}, \bibinfo {author} {\bibfnamefont {L.}~\bibnamefont {Pfeiffer}},\ and\ \bibinfo {author} {\bibfnamefont {K.}~\bibnamefont {West}},\ }\bibfield  {title} {\bibinfo {title} {Slow {R}eflection and {T}wo-photon {G}eneration of {M}icrocavity {E}xciton--polaritons},\ }\href@noop {} {\bibfield  {journal} {\bibinfo  {journal} {Optica}\ }\textbf {\bibinfo {volume} {2}},\ \bibinfo {pages} {1} (\bibinfo {year} {2015})}\BibitemShut {NoStop}%
\bibitem [{\citenamefont {Sun}\ \emph {et~al.}(2017)\citenamefont {Sun}, \citenamefont {Wen}, \citenamefont {Yoon}, \citenamefont {Liu}, \citenamefont {Steger}, \citenamefont {Pfeiffer}, \citenamefont {West}, \citenamefont {Snoke},\ and\ \citenamefont {Nelson}}]{sun2017bose}%
  \BibitemOpen
  \bibfield  {author} {\bibinfo {author} {\bibfnamefont {Y.}~\bibnamefont {Sun}}, \bibinfo {author} {\bibfnamefont {P.}~\bibnamefont {Wen}}, \bibinfo {author} {\bibfnamefont {Y.}~\bibnamefont {Yoon}}, \bibinfo {author} {\bibfnamefont {G.}~\bibnamefont {Liu}}, \bibinfo {author} {\bibfnamefont {M.}~\bibnamefont {Steger}}, \bibinfo {author} {\bibfnamefont {L.~N.}\ \bibnamefont {Pfeiffer}}, \bibinfo {author} {\bibfnamefont {K.}~\bibnamefont {West}}, \bibinfo {author} {\bibfnamefont {D.~W.}\ \bibnamefont {Snoke}},\ and\ \bibinfo {author} {\bibfnamefont {K.~A.}\ \bibnamefont {Nelson}},\ }\bibfield  {title} {\bibinfo {title} {Bose-{E}instein {C}ondensation of {L}ong-lifetime {P}olaritons in {T}hermal {E}quilibrium},\ }\href@noop {} {\bibfield  {journal} {\bibinfo  {journal} {Physical review letters}\ }\textbf {\bibinfo {volume} {118}},\ \bibinfo {pages} {016602} (\bibinfo {year} {2017})}\BibitemShut {NoStop}%
\bibitem [{\citenamefont {Caputo}\ \emph {et~al.}(2018)\citenamefont {Caputo}, \citenamefont {Ballarini}, \citenamefont {Dagvadorj}, \citenamefont {S{\'a}nchez~Mu{\~n}oz}, \citenamefont {De~Giorgi}, \citenamefont {Dominici}, \citenamefont {West}, \citenamefont {Pfeiffer}, \citenamefont {Gigli}, \citenamefont {Laussy} \emph {et~al.}}]{caputo2018topological}%
  \BibitemOpen
  \bibfield  {author} {\bibinfo {author} {\bibfnamefont {D.}~\bibnamefont {Caputo}}, \bibinfo {author} {\bibfnamefont {D.}~\bibnamefont {Ballarini}}, \bibinfo {author} {\bibfnamefont {G.}~\bibnamefont {Dagvadorj}}, \bibinfo {author} {\bibfnamefont {C.}~\bibnamefont {S{\'a}nchez~Mu{\~n}oz}}, \bibinfo {author} {\bibfnamefont {M.}~\bibnamefont {De~Giorgi}}, \bibinfo {author} {\bibfnamefont {L.}~\bibnamefont {Dominici}}, \bibinfo {author} {\bibfnamefont {K.}~\bibnamefont {West}}, \bibinfo {author} {\bibfnamefont {L.~N.}\ \bibnamefont {Pfeiffer}}, \bibinfo {author} {\bibfnamefont {G.}~\bibnamefont {Gigli}}, \bibinfo {author} {\bibfnamefont {F.~P.}\ \bibnamefont {Laussy}}, \emph {et~al.},\ }\bibfield  {title} {\bibinfo {title} {Topological {O}rder and {T}hermal {E}quilibrium in {P}olariton {C}ondensates},\ }\href@noop {} {\bibfield  {journal} {\bibinfo  {journal} {Nature materials}\ }\textbf {\bibinfo {volume} {17}},\ \bibinfo {pages} {145} (\bibinfo {year} {2018})}\BibitemShut {NoStop}%
\bibitem [{\citenamefont {Chomaz}\ \emph {et~al.}(2015)\citenamefont {Chomaz}, \citenamefont {Corman}, \citenamefont {Bienaim{\'e}}, \citenamefont {Desbuquois}, \citenamefont {Weitenberg}, \citenamefont {Nascimbene}, \citenamefont {Beugnon},\ and\ \citenamefont {Dalibard}}]{chomaz2015emergence}%
  \BibitemOpen
  \bibfield  {author} {\bibinfo {author} {\bibfnamefont {L.}~\bibnamefont {Chomaz}}, \bibinfo {author} {\bibfnamefont {L.}~\bibnamefont {Corman}}, \bibinfo {author} {\bibfnamefont {T.}~\bibnamefont {Bienaim{\'e}}}, \bibinfo {author} {\bibfnamefont {R.}~\bibnamefont {Desbuquois}}, \bibinfo {author} {\bibfnamefont {C.}~\bibnamefont {Weitenberg}}, \bibinfo {author} {\bibfnamefont {S.}~\bibnamefont {Nascimbene}}, \bibinfo {author} {\bibfnamefont {J.}~\bibnamefont {Beugnon}},\ and\ \bibinfo {author} {\bibfnamefont {J.}~\bibnamefont {Dalibard}},\ }\bibfield  {title} {\bibinfo {title} {Emergence of {C}oherence via {T}ransverse {C}ondensation in a {U}niform {Q}uasi-two-dimensional {B}ose {G}as},\ }\href@noop {} {\bibfield  {journal} {\bibinfo  {journal} {Nature communications}\ }\textbf {\bibinfo {volume} {6}},\ \bibinfo {pages} {6162} (\bibinfo {year} {2015})}\BibitemShut {NoStop}%
\bibitem [{\citenamefont {Malpuech}\ \emph {et~al.}(2003)\citenamefont {Malpuech}, \citenamefont {Rubo}, \citenamefont {Laussy}, \citenamefont {Bigenwald},\ and\ \citenamefont {Kavokin}}]{malpuech2003polariton}%
  \BibitemOpen
  \bibfield  {author} {\bibinfo {author} {\bibfnamefont {G.}~\bibnamefont {Malpuech}}, \bibinfo {author} {\bibfnamefont {Y.}~\bibnamefont {Rubo}}, \bibinfo {author} {\bibfnamefont {F.}~\bibnamefont {Laussy}}, \bibinfo {author} {\bibfnamefont {P.}~\bibnamefont {Bigenwald}},\ and\ \bibinfo {author} {\bibfnamefont {A.}~\bibnamefont {Kavokin}},\ }\bibfield  {title} {\bibinfo {title} {Polariton laser: thermodynamics and quantum kinetic theory},\ }\href@noop {} {\bibfield  {journal} {\bibinfo  {journal} {Semiconductor science and technology}\ }\textbf {\bibinfo {volume} {18}},\ \bibinfo {pages} {S395} (\bibinfo {year} {2003})}\BibitemShut {NoStop}%
\bibitem [{\citenamefont {Doan}\ \emph {et~al.}(2005)\citenamefont {Doan}, \citenamefont {Cao}, \citenamefont {Thoai},\ and\ \citenamefont {Haug}}]{doan2005condensation}%
  \BibitemOpen
  \bibfield  {author} {\bibinfo {author} {\bibfnamefont {T.}~\bibnamefont {Doan}}, \bibinfo {author} {\bibfnamefont {H.~T.}\ \bibnamefont {Cao}}, \bibinfo {author} {\bibfnamefont {D.~T.}\ \bibnamefont {Thoai}},\ and\ \bibinfo {author} {\bibfnamefont {H.}~\bibnamefont {Haug}},\ }\bibfield  {title} {\bibinfo {title} {Condensation kinetics of microcavity polaritons with scattering by phonons and polaritons},\ }\href@noop {} {\bibfield  {journal} {\bibinfo  {journal} {Physical Review B}\ }\textbf {\bibinfo {volume} {72}},\ \bibinfo {pages} {085301} (\bibinfo {year} {2005})}\BibitemShut {NoStop}%
\bibitem [{\citenamefont {Prokof’ev}\ and\ \citenamefont {Svistunov}(2002)}]{prokof2002two}%
  \BibitemOpen
  \bibfield  {author} {\bibinfo {author} {\bibfnamefont {N.}~\bibnamefont {Prokof’ev}}\ and\ \bibinfo {author} {\bibfnamefont {B.}~\bibnamefont {Svistunov}},\ }\bibfield  {title} {\bibinfo {title} {Two-dimensional weakly interacting bose gas in the fluctuation region},\ }\href@noop {} {\bibfield  {journal} {\bibinfo  {journal} {Physical Review A}\ }\textbf {\bibinfo {volume} {66}},\ \bibinfo {pages} {043608} (\bibinfo {year} {2002})}\BibitemShut {NoStop}%
\bibitem [{\citenamefont {Snoke}\ \emph {et~al.}(2023)\citenamefont {Snoke}, \citenamefont {Hartwell}, \citenamefont {Beaumariage}, \citenamefont {Mukherjee}, \citenamefont {Yoon}, \citenamefont {Myers}, \citenamefont {Steger}, \citenamefont {Sun}, \citenamefont {Nelson},\ and\ \citenamefont {Pfeiffer}}]{snoke2023reanalysis}%
  \BibitemOpen
  \bibfield  {author} {\bibinfo {author} {\bibfnamefont {D.}~\bibnamefont {Snoke}}, \bibinfo {author} {\bibfnamefont {V.}~\bibnamefont {Hartwell}}, \bibinfo {author} {\bibfnamefont {J.}~\bibnamefont {Beaumariage}}, \bibinfo {author} {\bibfnamefont {S.}~\bibnamefont {Mukherjee}}, \bibinfo {author} {\bibfnamefont {Y.}~\bibnamefont {Yoon}}, \bibinfo {author} {\bibfnamefont {D.}~\bibnamefont {Myers}}, \bibinfo {author} {\bibfnamefont {M.}~\bibnamefont {Steger}}, \bibinfo {author} {\bibfnamefont {Z.}~\bibnamefont {Sun}}, \bibinfo {author} {\bibfnamefont {K.}~\bibnamefont {Nelson}},\ and\ \bibinfo {author} {\bibfnamefont {L.}~\bibnamefont {Pfeiffer}},\ }\bibfield  {title} {\bibinfo {title} {Reanalysis of {E}xperimental {D}eterminations of {P}olariton-polariton {I}nteractions in {M}icrocavities},\ }\href@noop {} {\bibfield  {journal} {\bibinfo  {journal} {Physical Review B}\ }\textbf {\bibinfo {volume} {107}},\ \bibinfo {pages} {165302} (\bibinfo {year} {2023})}\BibitemShut {NoStop}%
\end{thebibliography}%
\clearpage
\title{Supplementary Information for: Coherence measurements of polaritons in thermal equilibrium reveal a new power law for two-dimensional condensates}
\date{\today}
\maketitle
\beginsupplement
\section{Sample Design and Experimental details}
We have repeated the experiment using two different samples grown at Princeton and at Waterloo, with similar design and cavity $Q$-factor. The results in the main text are all obtained from the Princeton sample. The two samples have identical design but with different quantum well thicknesses. The Princeton sample has a quantum well thickness of 7 nm, while the Waterloo sample has a quantum well thickness of 8.8 nm. We have designed both sample such that there is a slowly varying cavity gradient allowing us to have multiple detunings across the sample. We chose a location on the Waterloo sample that has the same detuning as the Princeton sample (near resonance) so that a direct comparison can be made. Importantly, the cavity gradient in these samples over short distances is negligible and therefore the polaritons do not feel a force because of this gradient.
\par
The two samples were pumped with different wavelengths since they had different thicknesses, resulting in different wavelength locations for the reflectivity minimum of the cavity. The Princeton sample was pumped with a laser tuned to the second reflectivity minimum (719.5 nm), about 113 meV above the lower polariton resonance. Similarly, the Waterloo sample was pumped with a laser tuned to the second reflectivity minimum (723.7 nm), about 144 meV above the lower polariton resonance.
\par
Since the two samples have a similar strcture and therefore a similar lifetime for the cavity photon, we observed evidence of thermalization in both samples. Typical examples for the occupation number as a function of energy for the Waterloo samples is shown in Fig. \ref{waterloo_thermalization}.
\begin{figure}
\includegraphics[width=0.9\columnwidth]{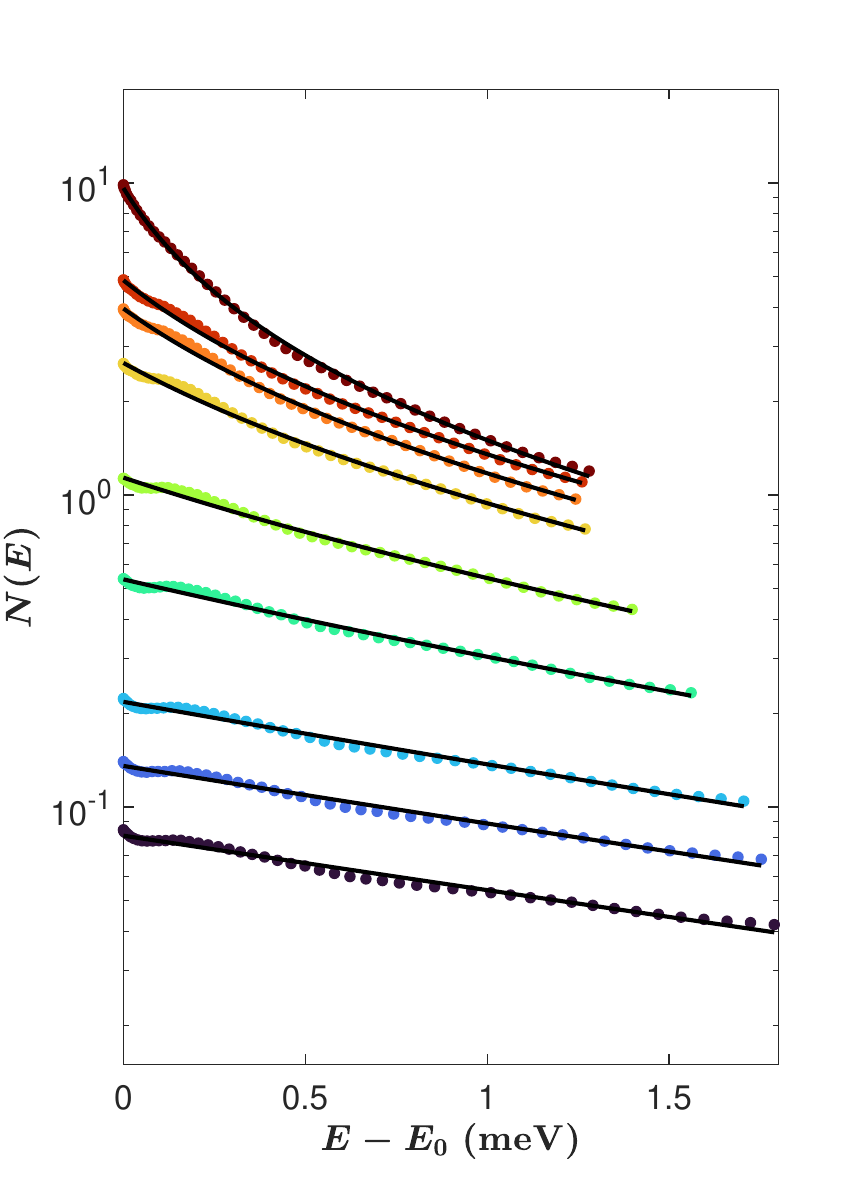}
\centering
\caption{\label{waterloo_thermalization} \textbf{Equilibrium distribution of polaritons.} Occupation of the lower polariton as a function of energy of the Waterloo sample. The solid lines are best fits to the equilibrium Bose-Einstein distribution.}
\end{figure}
\section{Density Calibration}
We used two different methods to calibrate the density of the polaritons and found consistency between them. In this section, we describe the procedure for each method. 
\par
In Fig.~5 of the paper, we have used the photon counting method to calibrate the density of the polaritons. To do this, we tuned the laser wavelength to 775 nm to match the polariton emission wavelength. This laser was then sent to a mirror at the sample plane. The mirror at the sample plane reflects the laser beam through the same optical path that was used in the experiment. We then imaged the reflected beam with the CCD camera. Since the total CCD count is proportional to the number of photons, the CCD counts can be written as:
\begin{equation}
\begin{aligned}
I_{\mathrm{CCD}} =\frac{1}{\eta}\frac{N_{\mathrm{ph}}}{\Delta t},
\end{aligned}
\end{equation}
where $\eta$ is the efficiency factor, $\Delta t$ is the integration time of the camera and $N_{\mathrm{ph}}$ is the number of photons detected by the camera. To find the number of photons sent to the camera, we measured the power of the laser. The number of photons that the camera detects during a time $\Delta t$ is then:
\begin{equation}
\begin{aligned}
N_{\mathrm{ph}} = \frac{P\Delta t}{hc/ \lambda},
\end{aligned}
\end{equation}
where $P$ is the measured power of the laser, $h$ is Planck constant, $c$ is the speed of light and $\lambda$ is the wavelength of the laser. The efficiency factor is then given by:
\begin{equation}
\begin{aligned}
\eta = \frac{P\lambda}{hc} \frac{1}{I_{\mathrm{CCD}}}.
\end{aligned}
\end{equation}
Therefore, this allows to relate the number of counts on the camera to the number of photons detected. We used this factor to calibrate the density of the polaritons. The total density of the polaritons for a given $I_{\mathrm{CCD}}$ count is then given by:
\begin{equation}
\begin{aligned}
n_{\mathrm{tot}} = \frac{\eta I_{\mathrm{CCD}}\tau}{A_{\mathrm{obs}}},
\end{aligned}
\end{equation}
where $\tau$ is the average radiative lifetime of the polaritons and $A_{\mathrm{obs}}$ is the observed area on the sample from which the light was collected. Since the excitation laser beam was chopped in the experiment with a duty cycle $d = 1.7\%$, then the total number of polaritons is 
\begin{equation}
\begin{aligned}
n_{\mathrm{tot}} = \frac{\eta I_{\mathrm{CCD}}\tau}{A_{\mathrm{obs}}d},
\end{aligned}
\end{equation}
\\
where $\tau \approx \tau_{\mathrm{cav}}/\left|C_{k_{\|}}\right|^2$. Here $\left|C_{k_{\|}}\right|^2$ is the photon fraction and $\tau_{\mathrm{cav}} = 135 \;\si{ps}$ \cite{steger2015slow} is the cavity lifetime. The observed area is given by the pinhole area $A_{\mathrm{obs}} = \pi (6\;\si{\mu m})^{2}$. 
\par
For the second method to calibrate the density of the polaritons, we make use of the fact that our system is in thermal equilibrium. To find the efficiency factor $\eta$, we minimized the mean-squared error in fitting a set of distributions $N(E)$ collected at different pump powers to the Bose–Einstein distribution with $T$ and $\mu$ as fit parameters for each distribution. This allows us to deduce one single collection efficiency that gives the best fit for distributions simultaneously. This means, for a set of $n$ number of distributions, we have a total of $2n+1$ free parameter, i.e $n$ temperature parameters, $n$ chemical potential parameters and one single efficiency factor. 
\par
We have found consistency between these two methods. At the threshold power $P/P_{\mathrm{th}} = 1$, the photon calibration methods predicts a density of $n = 3.20\;\si{\mu m^{-2}}$ while the best fit to the Bose-Einstein distribution in which
both $\mu$ and $T$ were allowed to vary predicts a density of $n = 3.12\;\si{\mu m^{-2}}$. 
\\
\section{Defining Critical Density}
To determine the critical density threshold of BEC, we measured the total emission intensity for different pump powers.  The threshold of condensation is defined from the “S'' curve in Fig.~\ref{scurve}. Near the condensation threshold, a nonlinear increase in intensity is observed, which becomes linear again at much higher pump power. This nonlinear increase in intensity is what we used to define the threshold of condensation.
\par
The procedure we used to define the power threshold and correspondingly, the density threshold is as follows: We first fitted the data in the linear regime with constant line in a log-log scale. We then defined the threshold at the point when the measured curve deviates from being linear by approximately $10\%$. Figure~\ref{scurve} shows the fit in the linear regime and the density threshold from the method discussed. The critical density we find is approximately $n_{th} = 3.2\;\si{\mu m^{-2}}$.
\par
The BKT transition density as a function of temperature has been calculated in Ref \cite{prokof2002two}, which is given by
\begin{equation}
\begin{aligned}
n_{c} = \frac{mk_{b}T}{2\pi\hbar^2}\ln\left ( \frac{\hbar^{2}\xi}{mg} \right ),
\end{aligned}
\end{equation}
where $\xi = 380$. For a temperature of $T = 20\;\mathrm{K}$, the calculated BKT transition density of our system is $n_{c} = 6.3\;\mathrm{\mu m^{-2}}$.
\begin{figure}
\includegraphics[width=0.9\columnwidth]{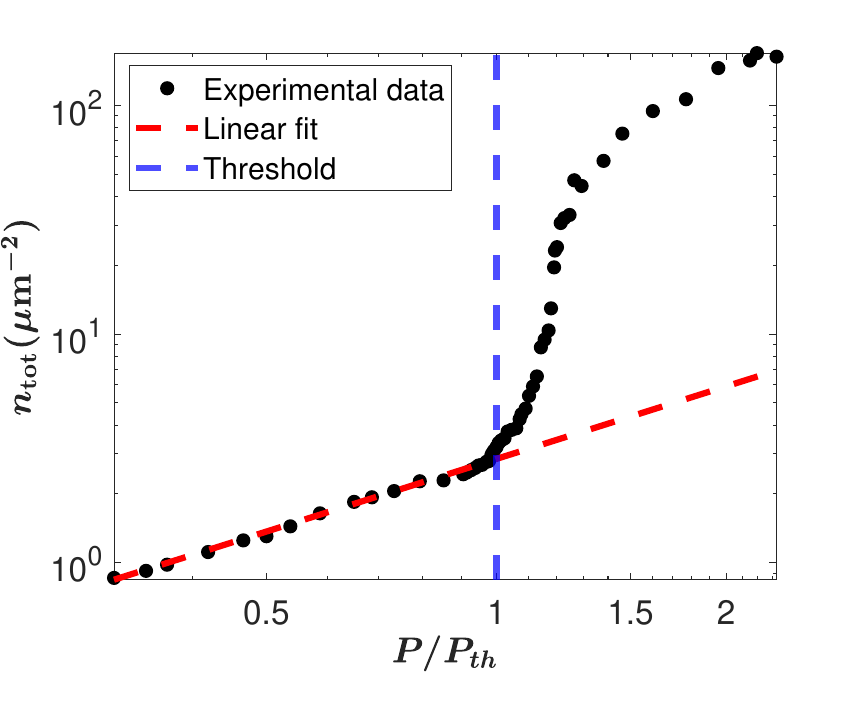}
\centering
\caption{\label{scurve} \textbf{Condensation threshold.} Polariton density as a function of pump power. The blue vertical line indicates the threshold, which is defined when the the measured curve deviates from being linear by approximately $10\%$. The threshold density is approximately $n_{th} = 3.2\;\si{\mu m^{-2}}$.}
\end{figure}
\section{Method of Extracting Occupation Number}
We have used angle-resolved imaging to measure the energy resolved emission in k-space $I(k,E)$  of the lower polariton. A typical $I(k,E)$ at low density is shown in Fig. \ref{E_k_image}\textbf{(A)}. To extract the occupation number $N(E)$ from the $I(k,E)$ image, we take vertical slices at each $k$ value to obtain $I(E)$ for each $k$ slice.  This $I(E)$ curve is then fit with a Lorentzian function to extract the polariton energy for each $k$ slice (see for example Fig.~\ref{E_k_image}\textbf{(B)}). The occupation for each $k$ slice is then given by:
\begin{equation}
\begin{aligned}
N(k_{i}) = C \tau(k_{i})\int \mathrm{d}E\; I(k_{i},E),
\end{aligned}
\end{equation}
where $\tau(k)$ is the $k$-dependent radiative lifetime, and $C$ is an overall constant that can be determined from photon counting. The value of $C$ is found by insuring that the total density for the extracted $N(E)$ curve, $\int \mathrm{d}E\; D(E) N(E)$ is equal to the density $n_{\mathrm{tot}}$ extracted from photon counting, that is 
\begin{equation}
\begin{aligned}
C = \frac{\int \mathrm{d}E\; D(E) N(E)}{n_{\mathrm{tot}}}.
\end{aligned}
\end{equation}
The same $C$ efficiency factor is then used to calibrate the density for each pump power and $D(E) = gm/2\pi\hbar^{2}$ is the density of states in two dimensions, with $g = 2$ to account for the spin degeneracy.
\par
We note that an alternate approach to find this efficiency factor is by using the second method described in the density calibration section. That is, to treat $C$ as a single efficiency factor that minimizes the mean-squared error in fitting a set of distributions $N(E)$ collected at different pump powers to the Bose–Einstein distribution.
\\
\begin{figure}
\includegraphics[width=0.9\columnwidth]{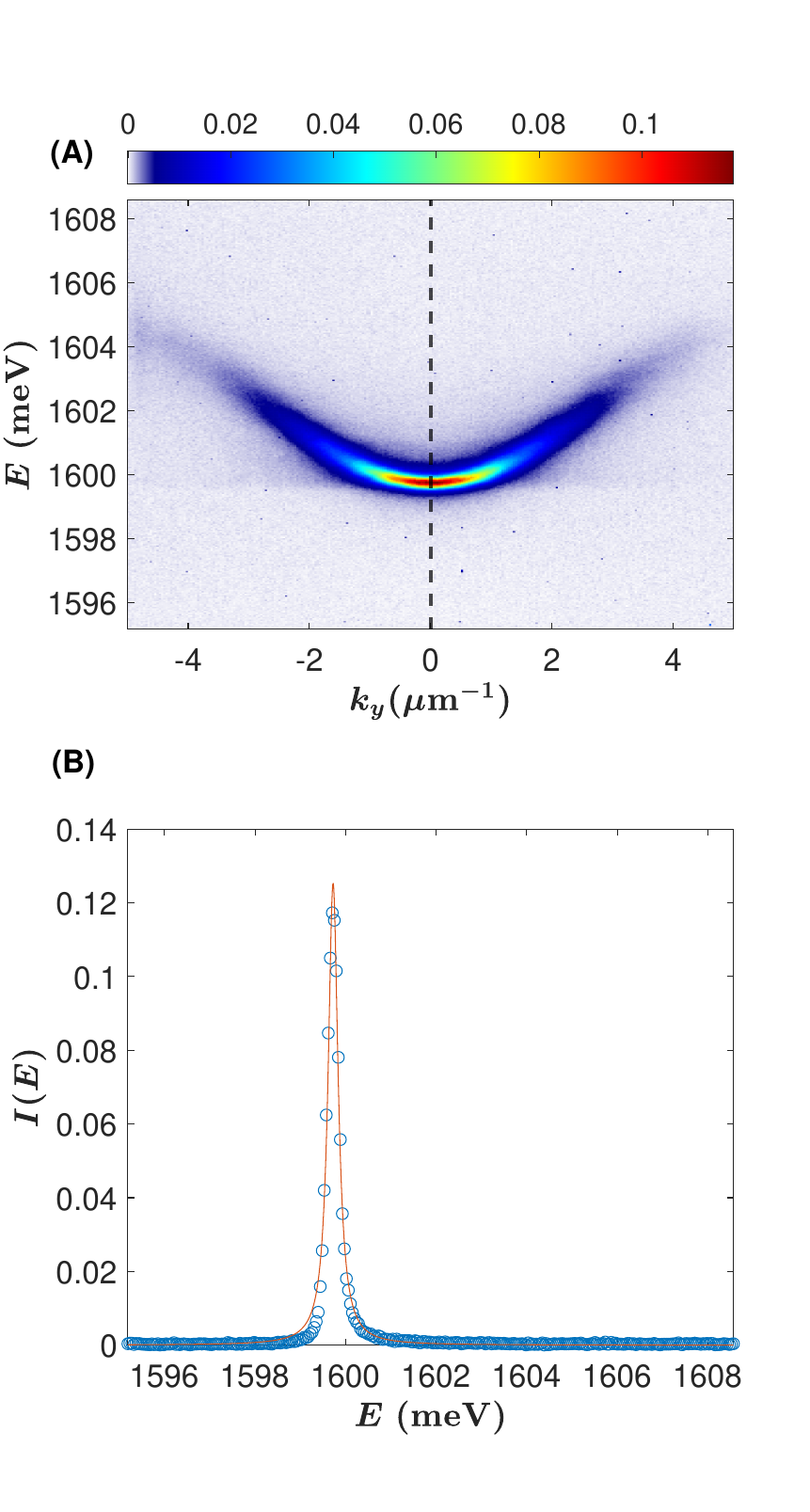}
\centering
\caption{\label{E_k_image} \textbf{Polariton occupation extraction method.} \textbf{(A)} A typical energy dispersion of the thermalized polariton gas. \textbf{(B)} A vertical slice in \textbf{(A)} at $k_{y} = 0$ showing the CCD counts. The red line is a fit to a Lorentzian function predicting an energy $1599.7$ meV for the polaritons at $k_{y} = 0$. The occupation number for this energy is proportional to the area under this Lorentzian curve.}
\end{figure}
\section{Energy Dispersion and Effective Mass}
Throughout the paper, we have assumed a parabolic dispersion with a fixed effective mass, namely in the density of state calculations and the G-P equation. Accounting for the full polariton dispersion should only
provide a small correction, because over the range of wave vectors that we measure, the polariton dispersion is very
parabolic (see Fig. \ref{parabolicity}). The maximum value of wave vector we can measure is determined by the
numerical aperture of the microscope objective, which allows a maximum cone of light to exit the lens. Since the
polariton dispersion is well approximated by a parabolic dispersion over the experimentally measured wave vector
range, we have used a fixed effective mass in the density of states calculations. Non-parabolic effects will only enter
for much higher wave vector, i.e. much higher temperature than the experimental conditions.
%%%%%%%%%%%%%%%%%%%%%%%%%%%%%%%%%%%%%%%%%%%%%%%%%%%%%%%%%%%%%%%%%%%%%%%%%%%%%%
\begin{figure*}
\includegraphics[width=0.9\textwidth]{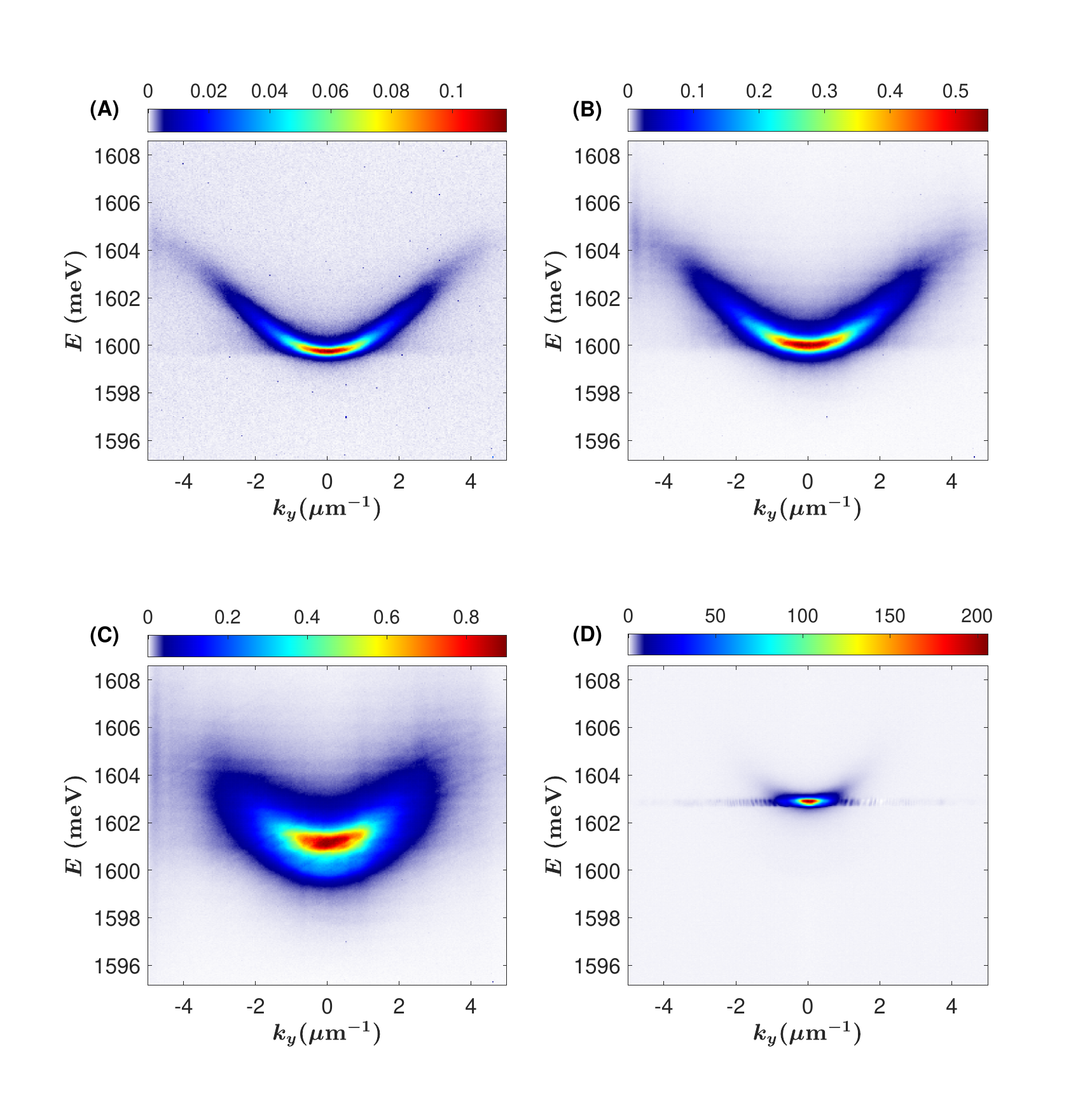}
\centering
\caption{\label{parabolicity}  \textbf{Energy dispersion of the lower polariton.} Typical energy dispersion images of the thermalized polariton gas at \textbf{(A)} $0.008\;P/P{_{\mathrm{th}}}$, \textbf{(B)}  $0.132\;P/P{_{\mathrm{th}}}$, \textbf{(C)}  $0.530\;P/P{_{\mathrm{th}}}$ and \textbf{(D)}  $ 1.265\;P/P{_{\mathrm{th}}}$ }
\end{figure*}
%%%%%%%%%%%%%%%%%%%%%%%%%%%%%%%%%%%%%%%%%%%%%%%%%%%%%%%%%%%%%%%%%%%%%%%%%%%%%%
\par
We note that many-body effects are expected to lead to renormalization of the mass, i.e density dependent mass. Figure \ref{mass_plot} shows the effective mass of the polaritons as a function of density. Over the range in which we measure the coherent fraction ($n_{\mathrm{tot}} >  1\; \mathrm{\mu m^{-2}}$), we find that the mass remains mostly constant as the density is increased. We cannot reliably extract the mass for densities larger than in Fig. \ref{mass_plot}. This is because at very high density, the condensate at $k=0$ becomes too bright and the occupation of states $k\neq 0$ are too dim that they are comparable to the noise in our measurements.
%%%%%%%%%%%%%%%%%%%%%%%%%%%%%%%%%%%%%%%%%%%%%%%%%%%%%%%%%%%%%%%%%%%%%%%%%%%%%%
\begin{figure}
\includegraphics[width=0.9\columnwidth]{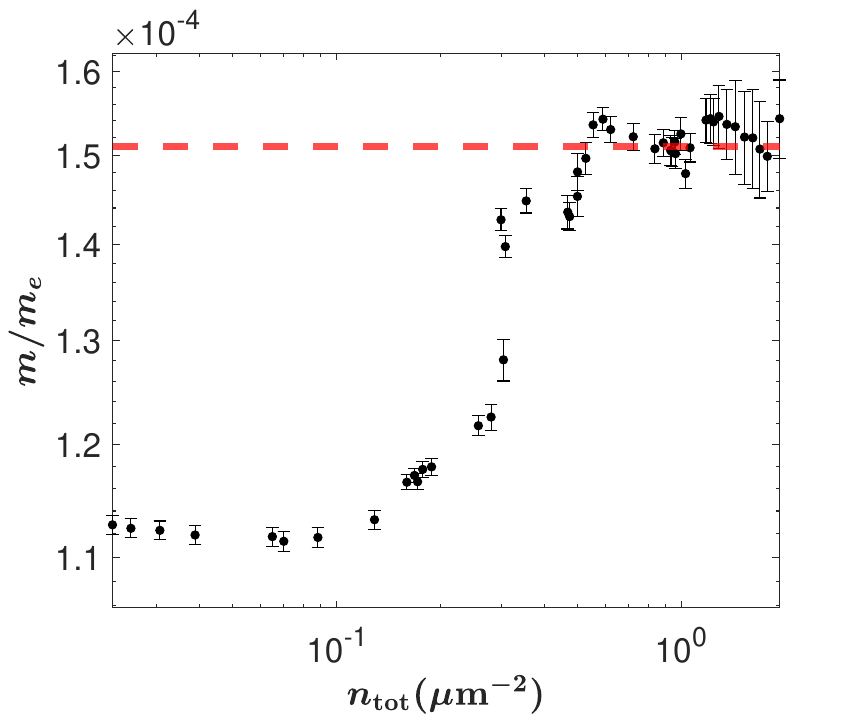}
\centering
\caption{\label{mass_plot} \textbf{Polariton effective mass.} The extracted lower polariton effective mass as a function of density. The dashed line is the mass used in the simulations.}
\end{figure}
%%%%%%%%%%%%%%%%%%%%%%%%%%%%%%%%%%%%%%%%%%%%%%%%%%%%%%%%%%%%%%%%%%%%%%%%%%%%%%
\section{Energy Linewidth}
In addition to extracting the occupation number from the fitting procedure described in the previous section, we have also extracted the energy linewidth $I(E)$ at $k=0$ for each pump power. The linewidth is extracted by fitting $I(E)$ at $k=0$ with a Lorentzian as shown in Fig.~\ref{E_k_image}\textbf{(B)} for each pump power. In this case, it is more helpful to plot the linewdith as a function of the occupation at ground state energy rather than the total polariton density. As shown in Fig.~\ref{linewidth}, the linewidth stays constant at low density and then increases due to collisional processes as the density increases. However, when the polartion gas becomes degenerate (i.e, $N(E=0)\sim 1$), significant line narrowing is observed, which is a hallmark of Bose-Einstein condensation. 
\begin{figure}
\includegraphics[width=0.9\columnwidth]{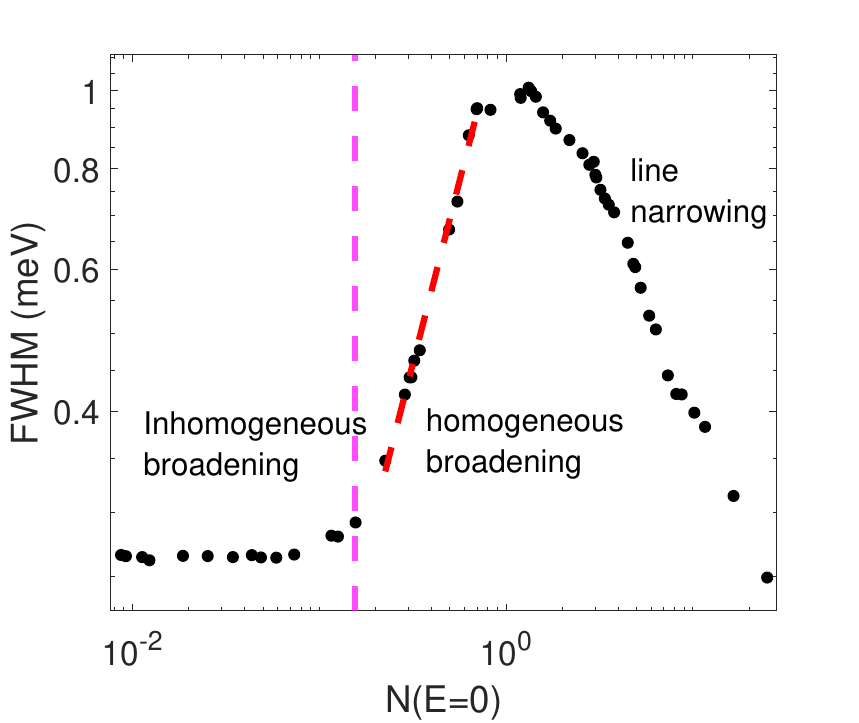}
\centering
\caption{\label{linewidth} \textbf{Linewidth narrowing and broadening.} Energy linewidth at $k=0$ as a function of the ground state occupation. Red line: power law of 1.0 indicating the regime of homogeneous broadening. Magenta line: a horizontal line indicating the regime of inhomogeneous broadening. Blue line:  The crossing regime between inhomogeneous broadening and homogeneous broadening. When the occupation becomes comparable to 1, the linewidth decreases sharply due to linewidth narrowing near the onest of condensation.}
\end{figure}
\section{Average Kientic Energy}
From the $T$ and $\mu$ fits shown in Fig.~3 of the main text, we have calculated the average energy per particle, 
\begin{equation}
\begin{aligned}
\bar{E} = \frac{\int \mathrm{d}E\; ED(E)N(E,T,\mu)}{\int \mathrm{d}E\;D(E)N(E,T,\mu)} =  \frac{\int \mathrm{d}E\; EN(E,T,\mu)}{\int \mathrm{d}E\; N(E,T,\mu)},\\
\label{eq:E_avg}
\end{aligned}
\end{equation}
where the density-of-states factor cancels out because the density of states is independent of energy in two dimensions. For each $T$ and $\mu$ values extracted from the fits, we plot the average energy per particle as a function of the total polariton density, as shown in Fig.~\ref{E_avg}\textbf{(A)}. Initially, the average energy per particle increases, presumably due to heating of the lattice due to the excess energy of the pump laser, seen also in the fit temperatures of Figure 3\textbf{(A)} of the main text. However, when the polariton gas becomes degenerate, the average energy per particle decreases significantly. 
\par
We take this decrease as primarily due to the fact that a degenerate Bose gas has lower average energy as the density increases, at constant temperature.
As a comparison, we calculate the average energy per particle expected for a Bose gas at a constant temperature (Fig.~\ref{E_avg}\textbf{(B)}). Since the occupation becomes peaked at $E = 0$ at high density, the average energy per particle has to decrease when the polariton gas becomes degenerate, which is in qualitative agreement with the experiment.   
\begin{figure}
\includegraphics[width=0.9\columnwidth]{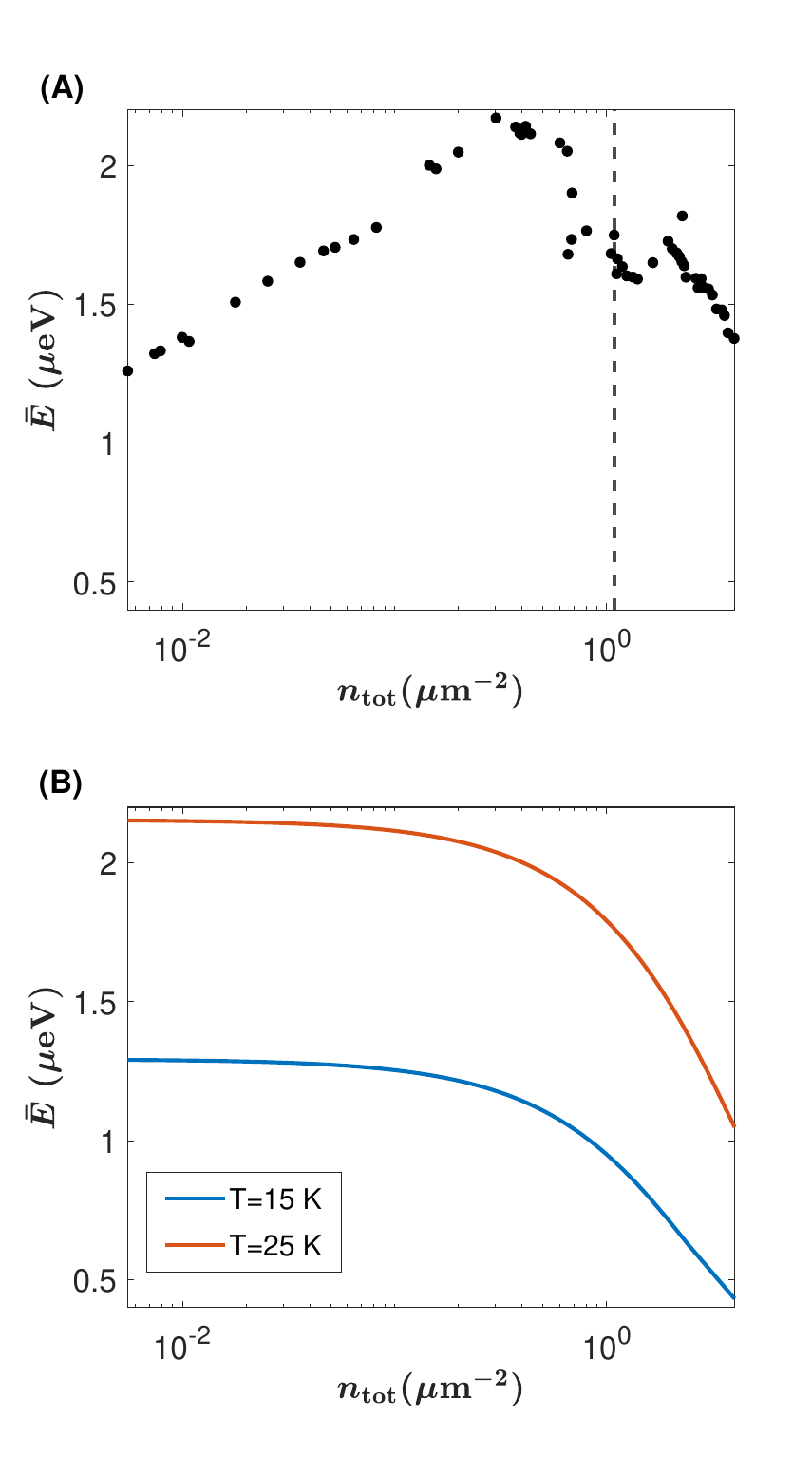}
\centering
\caption{\label{E_avg} \textbf{Average energy per particle.} \textbf{(A)} The experimental average energy per particle calculated from Eq.~\eqref{eq:E_avg} \textbf{(B)} The average energy per particle calculated using the Bose-Einstein distribution for a constant temperature $T = 15\;\si{K}$ and $T = 25\;\si{K}$.}
\end{figure}
\section{Inhomogeneity at High Density}
Although the polariton gas is fairly homogeneous for a large range of density within the observed area, at very high density we see evidence of self-trapping into the central region of the pump. To characterize the homogeneity of the polariton gas, we plot the full width at $90\%$ of the maximum of the polariton density extracted from the real space images as shown in Fig.~\ref{width}\textbf{(A)}. As a reference, we show at which density the polariton gas has a full width at $90\%$ of the maximum equal to the diameter of the pinhole in the experiment. This is shown in the $n_{0}/n_{\mathrm{tot}}$ plot in Fig.~\ref{width}\textbf{(B)} for the case of a pinhole with a radius $r = 6\; \mathrm{\mu m}$. 
\par
As seen in Fig.~\ref{width}\textbf{(B)}, the gas is fairly homogeneous for more than three orders of magnitude of the value of the coherent fraction. However, at very high density, the full width at $90\%$ maximum of the gas becomes smaller than the size of the pinhole leading to inhomogeneity of the polarion gas. 
\begin{figure}
\includegraphics[width=0.9\columnwidth]{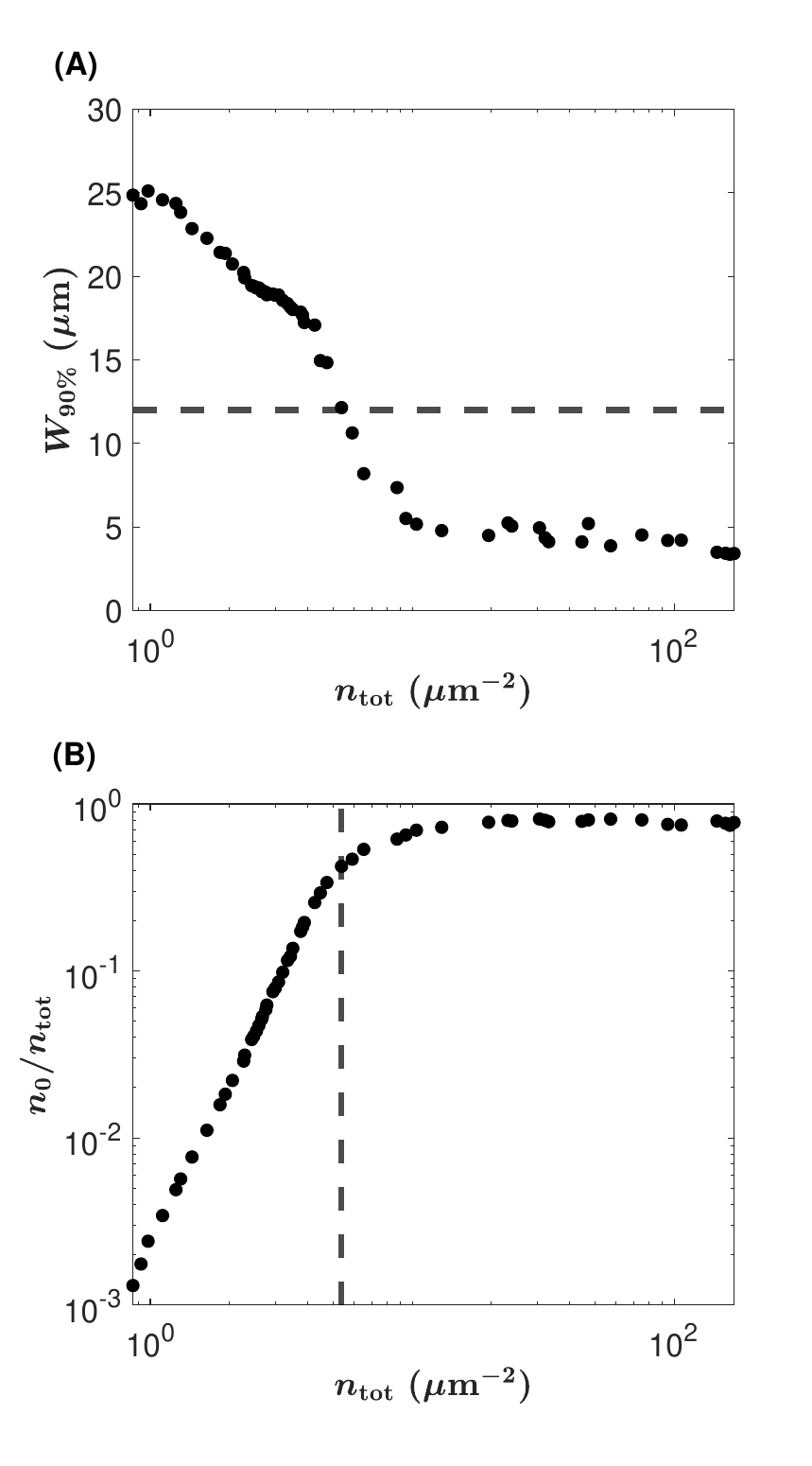}
\centering
\caption{\label{width} \textbf{Homogeneity of the polariton gas.} \textbf{(A)} Full width at $90\%$ maximum of the polariton density in real space as a function of the total integrated density. The dashed line indicates the diameter of the pinhole used in the main text. \textbf{(B)} The coherent fraction as a function of the total polariton density, for the same data set. The dashed line is the density at which the full width at $90\%$ maximum of the polariton profile becomes equal to the size of the pinhole.}
\end{figure}
\section{Polarization measurements}
%%%%%%%%%%%%%%%%%
\begin{figure*}
\includegraphics[width=0.9\textwidth]{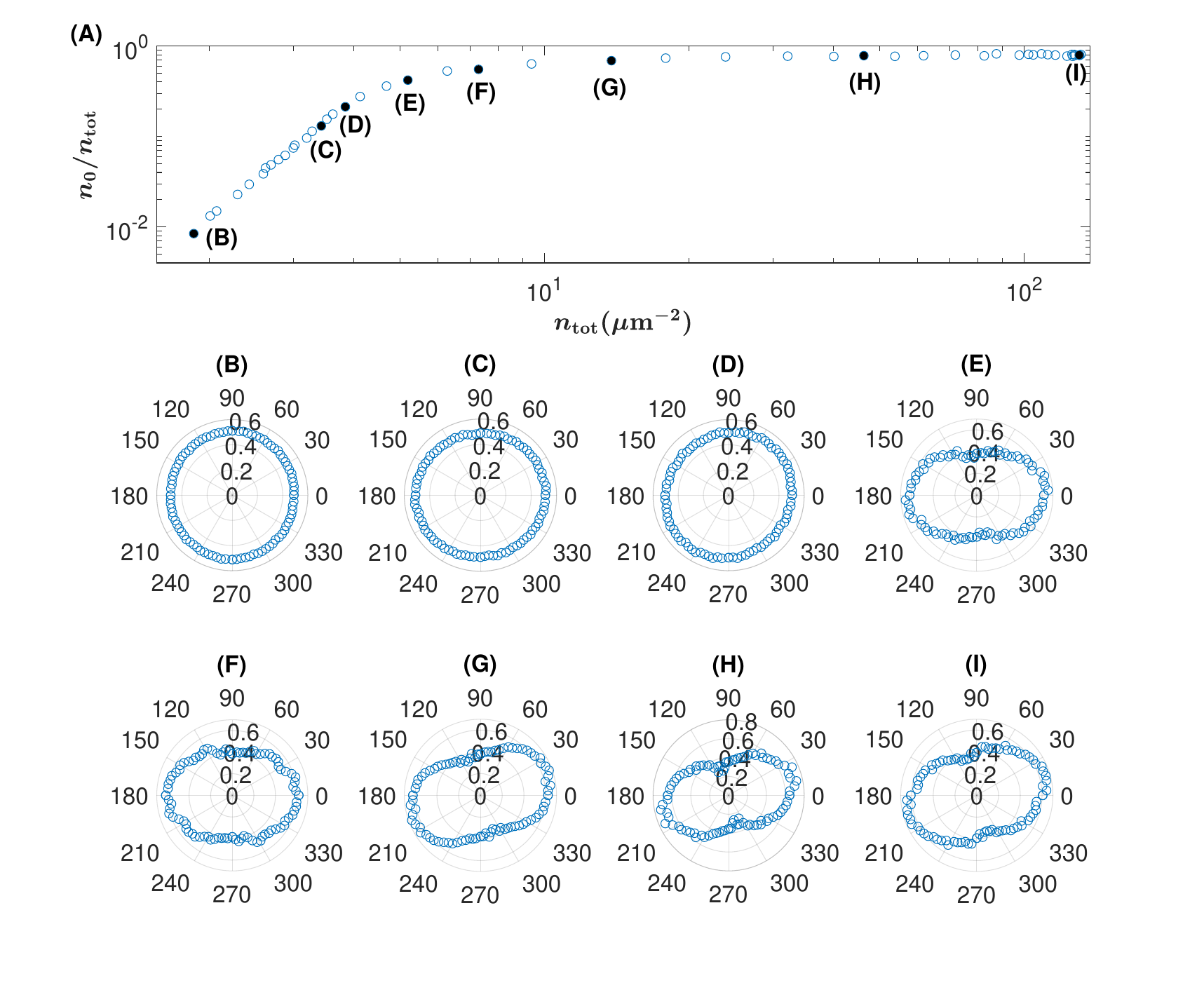}
\centering
\caption{\label{polaization} \textbf{Polarization of the polariton gas.} \textbf{(A)} Experimentally measured coherent fraction as a function of the total polariton density for a pinhole with an area $A = \pi (6\;\si{\mu m})^{2}$ \textbf{(B-I)} The intensity of the polariton gas as a function of the polarization angle in degrees. The points at which the polarization data are taken are labeled in \textbf{(A)} by the black solid circles. The polarization of the polariton gas at high density is mostly pinned to the $[110]$ crystalline axis, which corresponds to $15^{\circ}$. The gradient of the cavity is alighned along the horizontal direction.}
\end{figure*}
%%%%%%%%%%%%%%%%
We have measured the polarization of the polariton gas for different pump powers by using a half wave plate. The half wave plate was rotated in small angle steps and the intensity was recorded for each angle. This allows us to plot the polarization of the polariton gas for different densities as shown in Fig.~\ref{polaization}\textbf{(B-I)}. The polariton gas remains unpolarized in the power law regime and becomes polarized at very high density. Initially, the polarization direction becomes pinned to the gradient direction ($\theta =0^{\circ}$) of the cavity as shown in Fig.~\ref{polaization}\textbf{(E-F)}. At very high density, the polarization is pinned to the $[110]$ crystalline axis, which corresponds to $\theta = 15^{\circ}$ in polarization plot (Fig.~\ref{polaization}\textbf{(B-I)}.)
\section{Method of Extracting the Coherent Fraction}
As discussed in the main text, we have use the interference images in $k$-space to extract the coherent fraction. These interference images were then fit to the function,
\begin{equation}
\begin{aligned}
I (k)  =N(k)\left [ 1+ \alpha e^{-k/\kappa}\cos\left ( \lambda k \right )\right ],
\label{eq:SI_interference_fit}
\end{aligned}
\end{equation}
where $\kappa$ is a fit parameter giving the region of coherence, $\alpha$ is a fit parameter ranging between 0 and 1 giving the degree of coherence, and $\lambda$ is the component associated with the fringe spacing. For $N(k)$, we have used a Gaussian function of the form $N(k) = Be^{-k^2/\sigma_{k}^{2}}$, where $B$ and $\sigma_{k}$ are fit parameters. Therefore, in total we have four different parameters to fit $\kappa$, $\alpha$, $B$ and $\sigma_{k}$. The fringe spacing parameter $\lambda$ can be extracted from the data directly.
\par
We have used the same method to extract the coherent fraction for the experiment and the simulations. First, we take a slice of the interference pattern $k_{y} = 0$, which we then fit to the model mentioned above. Figure~\ref{coherence_fit} shows an example of the interference pattern $I(k_{y} = 0,k_{x})$ for the experiment and the resulting fit to Eq.~\eqref{eq:SI_interference_fit}. From this, we extracted the coherent fraction by computing the integral,
\begin{equation}
\begin{aligned}
\frac{n_{0}}{n_{tot}}=\frac{\alpha\int \mathrm{d}^{2}k\; N(k)e^{-k/\kappa}}{ \int \mathrm{d}^{2}k\; N(k)}.
\label{eq:SI_coherent_fraction}
\end{aligned}
\end{equation}
For example, for the interference patten in (Fig.~\ref{coherence_fit}\textbf{(B)}), we obtained the following parameters for the best fit, $\kappa = 0.92\; \si{\mu m}$, $\alpha = 0.96$, $B = 0.49$ and $\sigma_{k} = 0.34\;\si{\mu m}$. This then gives a coherent fraction $n_{0}/n_{\mathrm{tot}} = 0.70$. This procedure is repeated for each density, allowing us to plot the coherent fraction as a function of the polariton density. In Fig.~\ref{coherence_fit}\textbf{(A)}, we include a typical fit to the interference pattern below the threshold.
\begin{figure*}
\includegraphics[width=0.9\textwidth]{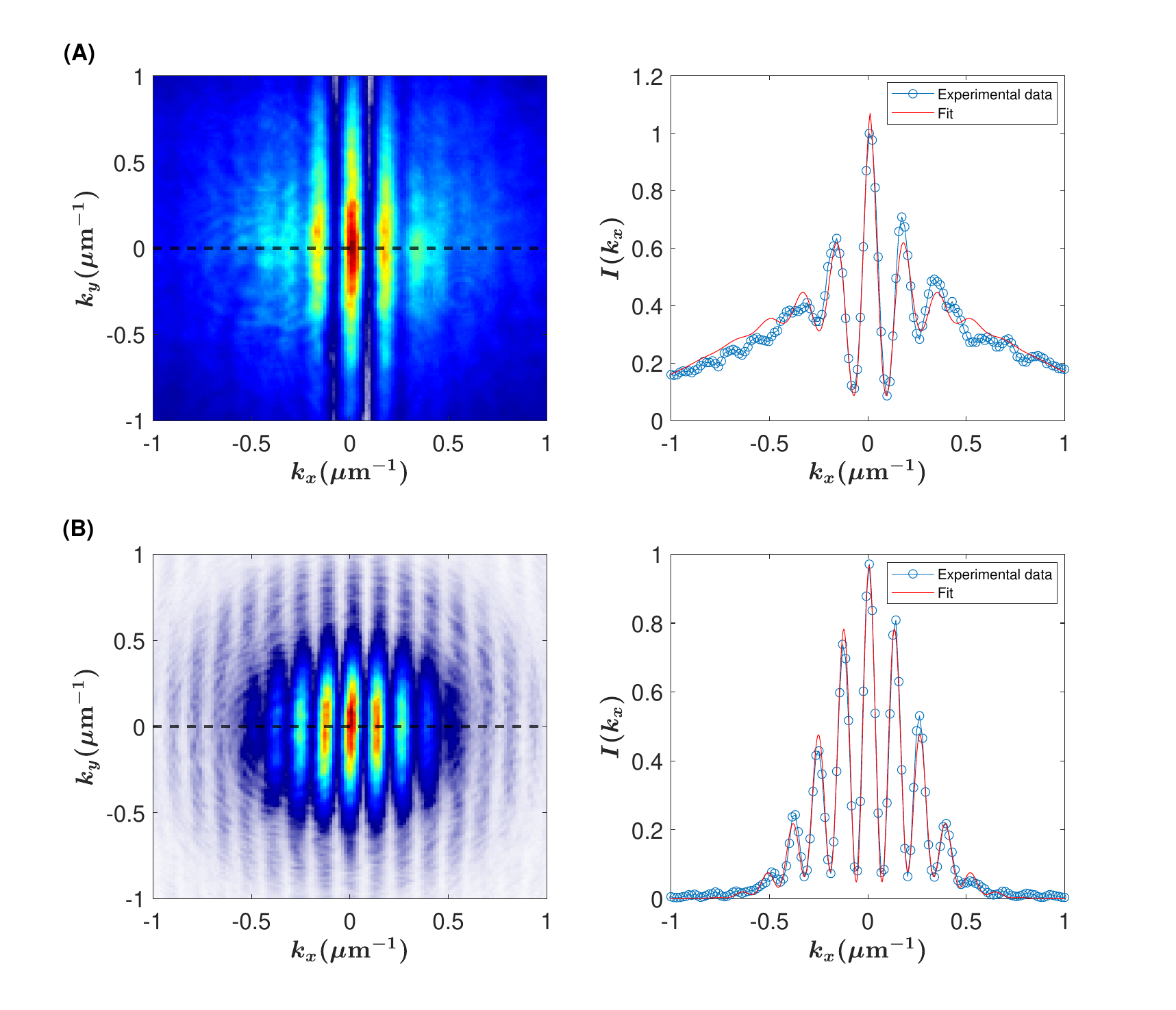}
\centering
\caption{\label{coherence_fit} \textbf{Low and high density examples of the best fit to Eq.~\eqref{eq:SI_interference_fit}.} \textbf{(A, left panel)} A typical experimental interference image in $k$-space below the threshold. \textbf{(A, right panel)} horizontal slice in \textbf{(A, left panel)} at $k_{y} = 0$ as indicated by the dashed line. \textbf{(B, left panel)} A typical experimental interference image in $k$-space above the threshold. \textbf{(B, right panel)} horizontal slice in \textbf{(B, left panel)} at $k_{y} = 0$ as indicated by the dashed line. The red line is the best fit to Eq.~\eqref{eq:SI_interference_fit}}
\end{figure*}
\section{Temporal Coherence Measurements}
In addition to measuring the coherence in $k$-space, we have also measured the coherence time of the polariton gas. By introducing a time delay between the two interferometer arms, we have measured the time correlation function $g^{(1)}(\Delta t)$. This is done by computing the integral of the visibility, that is $n_{0}/n_{tot}$, using the procedure described in the previous section (Eq.~\eqref{eq:SI_coherent_fraction}) for each time delay as shown in Fig.~\ref{coherence_time}\textbf{(A)}. To extract the coherence time, we have fitted the visibility function with a Gaussian (Fig.~\ref{coherence_time}\textbf{(A)}), allowing us to extract the coherence time for each density. The coherence time is defined as the FWHM of the fitted Gaussian. Figure~\ref{coherence_time}\textbf{(B)} shows the coherence time as a function of the polarion density. The value of about 2 ps at the lowest density is consistent with the line broadening reported at that density in previous work, which was attributed to decoherence due to polariton-exciton collisions \cite{snoke2023reanalysis}. 
\begin{figure}
\includegraphics[width=0.9\columnwidth]{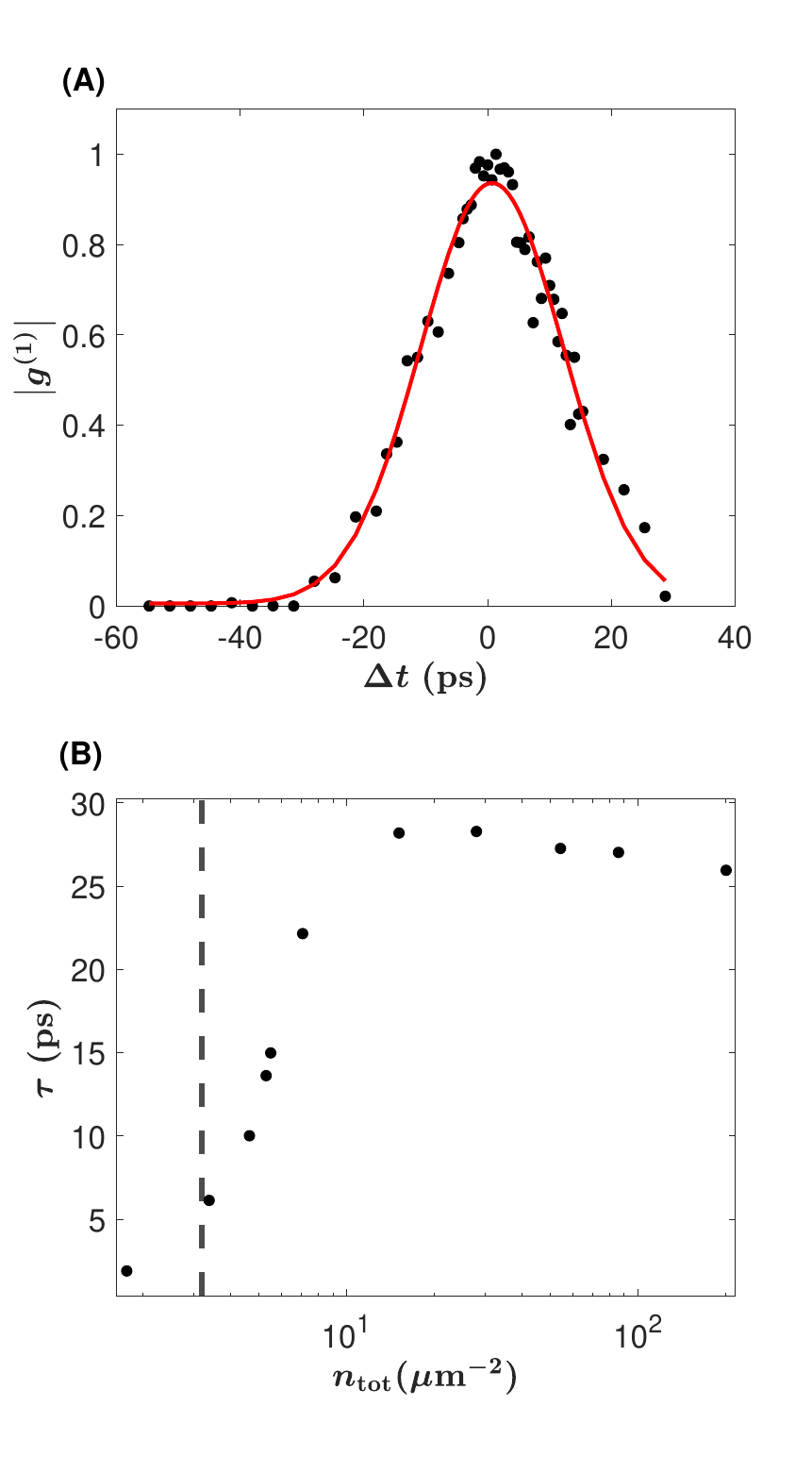}
\centering
\caption{\label{coherence_time} \textbf{Coherence time}. \textbf{(A)} A typical visibility curve as a function of the time delay between the two Michelson arms. The red line is a Gaussian fit used to extract the coherence time. \textbf{(B)} Coherence time as a function of density measured by varying the arm of the Michelson interferometer. The vertical line indicates the critical density.}
\end{figure}
\section{Effect of changing pinhole size}
We have found that the same power law for the coherent fraction is seen for different pinhole sizes from which the light is collected in real space. The largest pinhole that was used experimentally has a radius of $6\; \si{\mu m}$ since for larger pinhole sizes, the assumption of homogeneity breaks down (see Fig~\ref{width}\textbf{(A)}). Figure~\ref{coherence_pinhole}\textbf{(A)} shows a comparison of the experimentally measured coherent fraction for three examples of different pinhole sizes. The same power law is observed for these different cases.
\begin{figure}
\includegraphics[width=.9\columnwidth]{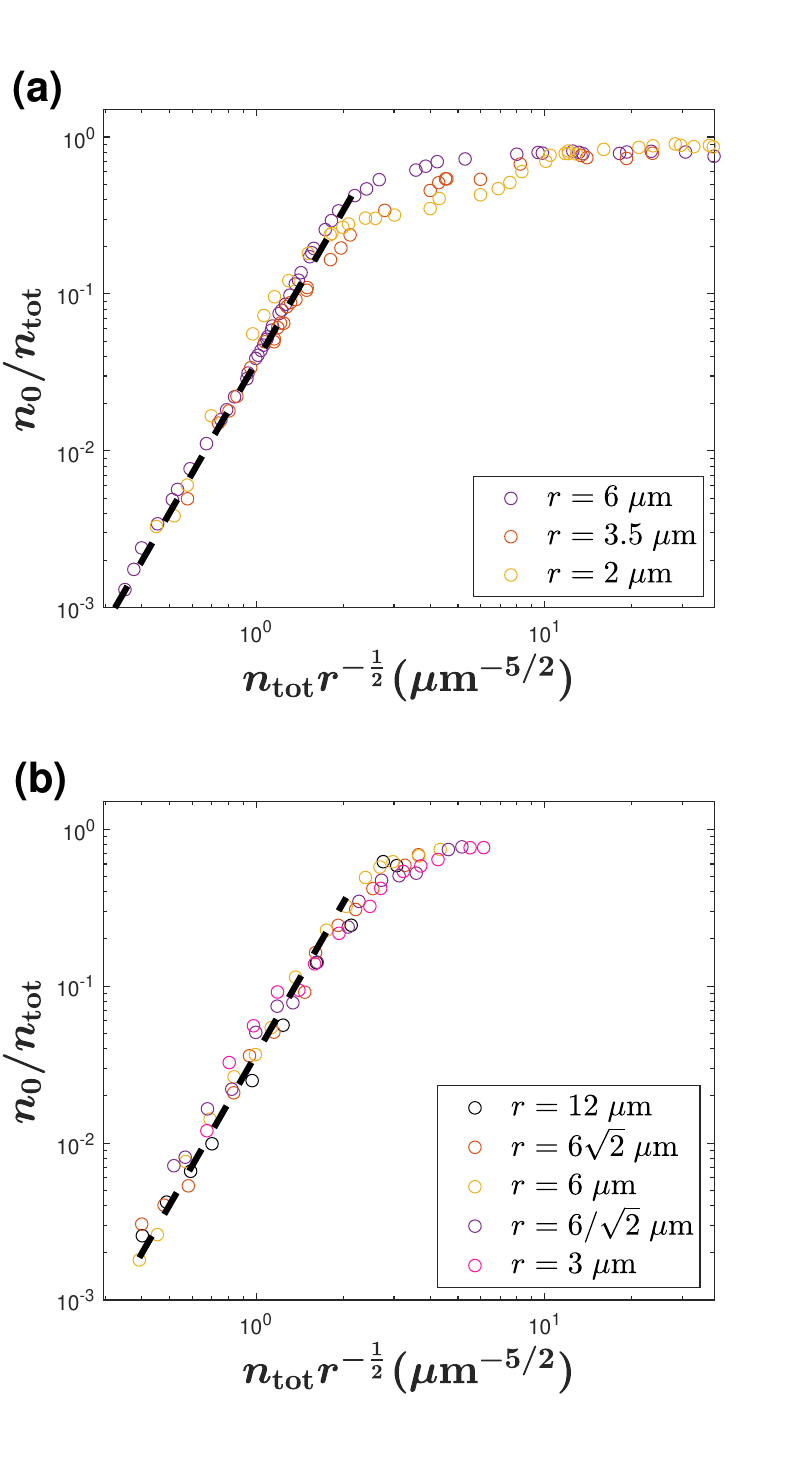}
\centering
\caption{\label{coherence_pinhole} \textbf{Effect of different pinhole sizes.} \textbf{(A)} The experimentally measured coherent fraction for three different pinhole sizes collapsed onto a single curve. \textbf{(B)}  The condensate fraction extracted from the numerics for five different pinhole sizes collapsed onto a single curve. The dashed line is a power law of 3.2.}
\end{figure}
\par
In agreement with the experiment, our numerical model shows that the effect of the aperture size gives a shifted curve with the same power law. Figure~\ref{coherence_pinhole}\textbf{(B)} shows five different examples, starting from a radius of $r = 3\;\si{\mu m}$ and increase the radius by a factor of $\sqrt{2}$, which corresponds to increasing the area by a factor of two. Although the largest area we can explore experimentally is $\pi( 6\;\si{\mu m})^{2}$, our numerical analysis shows that this power law is still maintained even after increasing the area by a factor of four. 
\par
\section{Real-space correlation function}
In addition to calculating the coherent fraction, we have also calculated the correlation length in our numerics. For a two-dimensional system, the correlation function can be written as
\begin{equation}
\begin{aligned}
g^{(1)}(\Delta r)=\frac{\left\langle\psi^*\left(r+\Delta r, t\right) \psi\left(r, t\right)\right\rangle}{\sqrt{\left\langle\left|\psi\left(r+\Delta r, t\right)\right|^2\right\rangle\left\langle\left|\psi\left(r, t\right)\right|^2\right\rangle}}.
\label{eq:correlation_function}
\end{aligned}
\end{equation}
The correlation function is calculated at a sufficiently late time after transients have died down.  Figure~\ref{correlation_function}\textbf{(A)} shows the correlation function calculated for different densities. A crossover from exponential to algebraic decay of the first order correlations is clearly observed, which is a characteristic feature of the BKT transition. 
\begin{figure}
\includegraphics[width=.9\columnwidth]{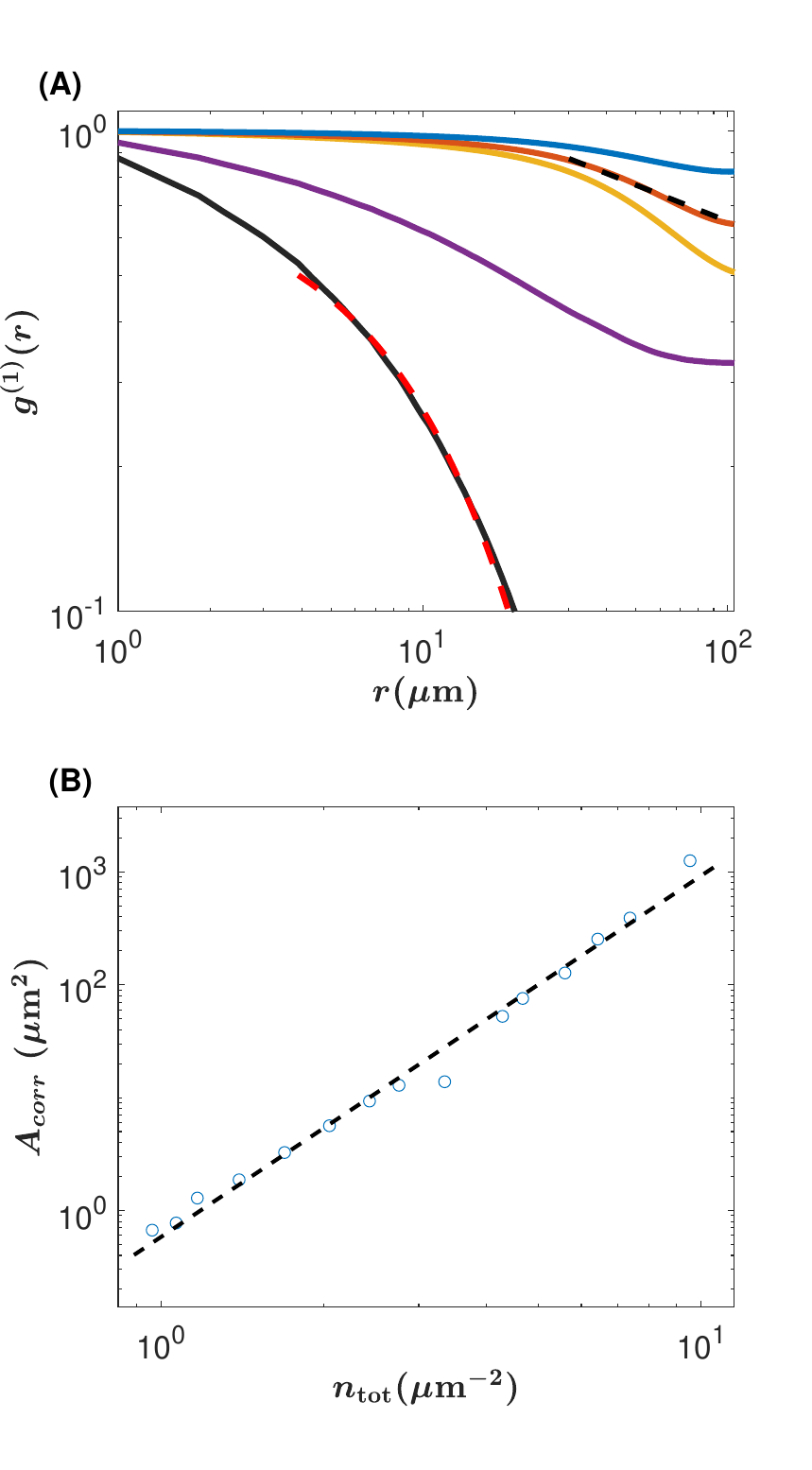}
\centering
\caption{\label{correlation_function} \textbf{Correlation function and correlation area.} \textbf{(A)} The correlation function calculated for density values from small to high, $ 7.29\;\si{\mu m^{-2}}$, $ 11.71\;\si{\mu m^{-2}}$, $ 14.97\;\si{\mu m^{-2}}$, $ 15.39\;\si{\mu m^{-2}}$ and $ 16.11\;\si{\mu m^{-2}}$. Red dashed line: fit to exponential decay. Black dashed line is a $r^{-1/4}$ power law, which is expected at the BKT transition density. \textbf{(B)} The correlation area minus the zero-density correlation area, as a function of the total density. The black dashed line is a power law of 3.2.}
\end{figure}
\section{Correlation area}
In this section, we make a connection between the coherent fraction and the correlation area. The interference pattern we measure in k-space can be written as:
\begin{equation}
\begin{split}
I(\vec{k}) =& \left | \psi\left (\vec{k}\right )+\psi\left (-\vec{k}\right )  \right |^{2}\\
=& \left | \psi\left (\vec{k}\right ) \right |^{2} + \left | \psi\left (-\vec{k}\right ) \right |^{2}+2\Re \left [ \psi^{*}\left (\vec{k}\right )\psi\left (-\vec{k}\right ) \right ].
\end{split}
\end{equation}
Assuming $ \left | \psi\left (\vec{k}\right ) \right |^{2} =  \left | \psi\left (-\vec{k}\right ) \right |^{2} = N(k)$ gives,
\begin{equation}
\begin{split}
I(\vec{k}) = 2N(k) +2\Re \left [ \psi^{*}\left (\vec{k}\right )\psi\left (-\vec{k}\right ) \right ].
\end{split}
\end{equation}
One, therefore, can define a coherent fraction $\Delta = n_{0}/n_{\mathrm{tot}}$ as:
\begin{equation}
\begin{aligned}
\Delta  = \frac{\int \mathrm{d}^{2}k\; \psi^{*}(\vec{k})\psi(-\vec{k})}{\int \mathrm{d}^{2}k\; \left | \psi(\vec{k}) \right |^{2}},
\end{aligned}
\label{eq:ntot_psi_k}
\end{equation}
where the wavefunction in $k$-space is given by the Fourier transform of $\psi(\vec{r})$.
\begin{equation}
\begin{aligned}
\psi(\vec{k}) = \frac{1}{2\pi}\int \mathrm{d}^{2}r \;\psi(\vec{r})e^{i\vec{k}\cdot \vec{r}}.
\end{aligned}
\label{eq:psik}
\end{equation}
Plugging in Eq.~\ref{eq:psik} into Eq.~\ref{eq:ntot_psi_k} and using the relation $\int \mathrm{d}^{2}k \; e^{-i\vec{k}\cdot \left ( \vec{r}+\vec{r'} \right )} = 2\pi \delta\left( \vec{r}+\vec{r'} \right )$, we obtain
\begin{equation}
\begin{aligned}
\Delta  = \frac{\int \mathrm{d}^{2}r\; \psi^{*}(\vec{r})\psi(-\vec{r})}{\int \mathrm{d}^{2}r\; \left | \psi(\vec{r}) \right |^{2}}.
\end{aligned}
\end{equation}
The origin $\vec{r} = 0$ is defined arbitrarily here. In a translationally invariant system, we should average over all origins $\vec{r}$:
\begin{equation}
\begin{aligned}
\Delta  &= \frac{1}{n_{\mathrm{tot}}A}\int \mathrm{d}^{2}r\; \left \langle \psi^{*}(\vec{r})\psi(-\vec{r}) \right \rangle\\
 &= \frac{1}{n_{\mathrm{tot}}A^{2}} \int \mathrm{d}^{2}r'\int \mathrm{d}^{2}r\;  \psi^{*}(\vec{r'}+\vec{r})\psi(\vec{r'}-\vec{r})\\
  &= \frac{1}{n_{\mathrm{tot}}A^{2}} \int \mathrm{d}^{2}r\int \mathrm{d}^{2}r''\; \psi^{*}(\vec{r''}+2\vec{r})\psi(\vec{r''}),
\end{aligned}
\end{equation}
where $\vec{r''} = \vec{r'}-\vec{r}$, $A$ is the area and $n_{\mathrm{tot}}$ is the total density. Using the definition of the correlation function in Eq.~\ref{eq:correlation_function}, we have:
\begin{equation}
\begin{aligned}
\begin{split}
\Delta &= \frac{1}{A} \int \mathrm{d}^{2}r\;g^{(1)}(2r)\\
&\equiv  \frac{A_{\mathrm{corr}}}{A},
\label{eq:coherent_fraction_A}
\end{split}
\end{aligned}
\end{equation}
where $A_{\mathrm{corr}}$ is the correlation area, which is given by
\begin{equation}
\begin{aligned}
A_{\mathrm{corr}} =\int \mathrm{d}^{2}r\;g(2r).
\label{eq:correlation_area}
\end{aligned}
\end{equation}
Therefore, in the low density limit, the coherent fraction is proportional to the correlation area $\Delta\sim A_{\mathrm{corr}}$. The correlation length is related to the correlation area via the relation
\begin{equation}
\begin{aligned}
l_{\mathrm{corr}} = \sqrt{\frac{A_{\mathrm{corr}} }{\pi}}\sim \Delta^{\frac{1}{2}}.
\label{eq:correlation_length}
\end{aligned}
\end{equation}
Figure~\ref{correlation_function}\textbf{(B)} shows the correlation area calculated from our model as a function of density. The correlation area is fit by a power law of $n^{3.17\pm 0.21}$. (Similar to the condensate fraction calculations, we subtract the zero-density limit of the correlation area, which corresponds to the Maxwell-Boltzmann limit.) This power law is consistent with the experimental measurements and with the calculated coherent fraction from the numerics since $A_{\mathrm{corr}}\sim \Delta \sim n^{3.2}$.  The power law is still maintained in the numerics even for the full system size of $L = 210\; \mathrm{\mu m}$, which is much larger than the thermal de Broglie wavelength of the polaritons $L >> \lambda_{th} = h/\sqrt{2\pi m k_{B}T}\sim 1~\mathrm{\mu m}$.
\section{Comparison of different data sets}
We have found consistent results for different data sets that were obtained during different experiments. We have also repeated the experiment using two different samples grown at Princeton and at Waterloo, with similar design and cavity $Q$-factor (i.e., comparable polariton lifetime) and have found consistent results, in particular, the same power law for the quasicondensate fraction. Figure~\ref{multiple_data_sets} compares multiple data sets. 
\begin{figure}
\includegraphics[width=0.9\columnwidth]{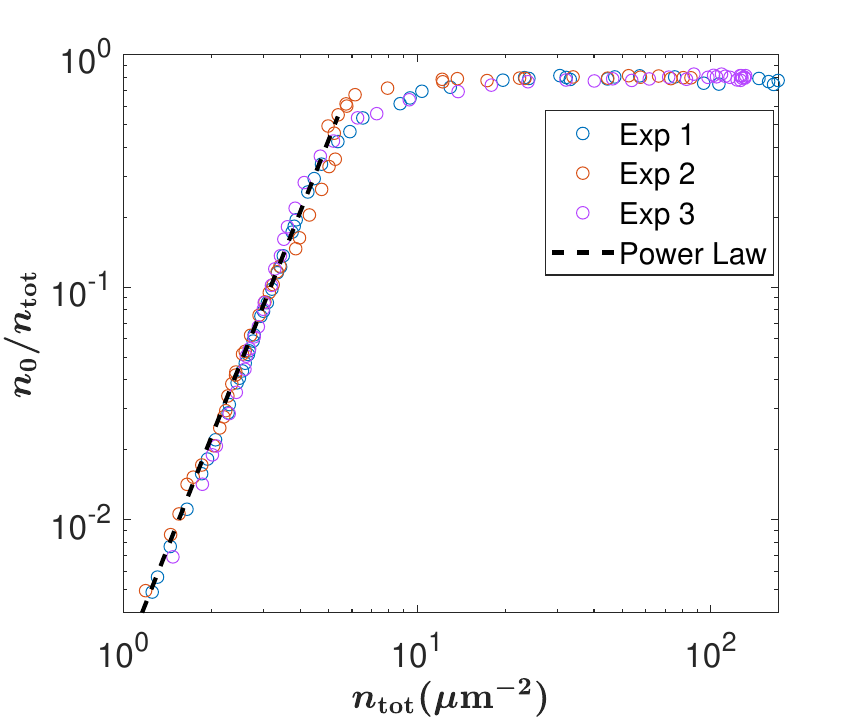}
\centering
\caption{\label{multiple_data_sets} \textbf{Multiple data sets for the coherent fraction.} Different data sets showing the same power law for a pinhole with an area $A = \pi (6\;\si{\mu m})^{2}$. Exp 1 is the results obtained from the Princeton sample. Exp 2 is obtained from the same sample on a different day and a different location on the sample. Exp 3 is obtained from the Waterloo sample. The dashed line is a power law of 3.2.}
\end{figure}

\section{Numerical Method}
As mentioned in the main text, we solved the Gross-Pitaevskii equation with noise introduced in the initial conditions, 
\begin{equation}
i \hbar \frac{\partial  \psi(\mathbf{r},t)}{\partial t}=\left[-\frac{\hbar^{2}\nabla^2 }{2m}+g_c|\psi(\mathbf{r},t)|^2\right] \psi(\mathbf{r,t}).
\end{equation}
The Gross-Pitaevskii is solved numerically adopting a Runge-Kutta method of order fourth on a two-dimensional numerical grid with $N=512^2$ points with grid-spacing $a = 0.41\; \si{\mu m}$. 
We have used the following experimental parameters in the fit shown: $m=1.515\times 10^{-4} m_e$ with $m_e$ the electron mass, and $g~= ~2.105~\si{\mu eV- \mu m^2}$ for the interaction strength. Because the correlation of the coherence depends sensitively on $gn$, where $n$ is the density, and we know the absolute density of the polaritons from the good fits to the Bose-Einstein distribution, as discussed in the main text, these experiments are an independent measurement of the polariton-polariton interaction constant. Accounting for the uncertainty in the density calibration, we find $g~= ~2.11\pm 0.25~\si{\mu eV- \mu m^2}$.  Since the polaritons are approximately 55\% excitonic in these experiments, this corresponds to a polariton-exciton interaction constant {$\tilde{g} = 3.83 ~\si{\mu eV- \mu m^2}$, or an exciton-exciton interaction constant of $6.96 ~\si{\mu eV- \mu m^2}$. The number found here is below some previously reported experimental values, but well above the theoretical prediction of $g \sim ~ 0.25 ~\si{\mu eV- \mu m^2}$ \cite{snoke2023reanalysis}.
% D. W. Snoke, V. Hartwell  J. Beaumariage, S. Mukherjee, Y. Yoon , D. M. Myers, M. Steger, Z. Sun , K. A. Nelson, and L. N. Pfeiffer, Reanalysis of experimental determinations of polariton-polariton interactions in microcavities, Phys. Rev. B {\bf 107}, 165302 (2023). 

To avoid reflections from the boundaries, we have used periodic boundary conditions. We have ensured that system size is sufficiently larger than the experimental size of the system to avoid any boundary effects influencing the relevant region. The numerical results presented in the main text are for a system size of $210\; \si{\mu m}$.  
\par
We have assumed that the system is initially given by an incoherent equilibrium state,
\begin{equation}
\begin{split}
\psi\left (x,y,t=0\right ) = \sum_{k_{n}}\sum_{k_{m}}\sqrt{ N\left (\sqrt{k^2_n+k^2_m} \right ) }\;e^{i\left ( k_{n}x+ k_{m}y\right )}\\
\times e^{i\left ( \theta_{k_{n}}+ \theta_{k_{m}}\right )},
\label{eq:SI_initial_conditions}
\end{split}
\end{equation}
where $N\left (k  \right )$ is given by the Bose-Einstein distribution,
\begin{equation}
\begin{split}
N\left (k  \right ) = \frac{1}{e^{\left (E(k)-\mu  \right )/k_{B}T} -1}.
\end{split}
\end{equation}
We have fixed the temperature to be $T = 20\;\si{K}$ in the numerics and varied the chemical potential $\mu$. Therefore, the choise of $\mu$ defines the density of the system, which is given by, 
\begin{equation}
\begin{aligned}
n_{\mathrm{tot}} = \int \mathrm{d}E\; D(E) N(E,T,\mu)
\end{aligned}
\end{equation}
where $D(E)$ is the constant density of states in two-dimensions. For each chemical potential value, the wavefunction was evolved in time for $t_{\mathrm{max}} = 1\;\si{ns}$. The interference pattern $I\left ( k_{x},k_{y} \right )$ was then calculated using the equation described in the main text and then averaged over $20$ independent stochastic paths each with different random initial conditions for $\theta_{k_{x}}$ and $\theta_{k_{y}}$.
\end{document}